\documentclass{emulateapj}
\usepackage{natbib,graphicx,url}
\slugcomment{\sc Accepted for Publication in ApJ}
\shorttitle{Radio-quiet Quasars with Weak Emission Lines}
\shortauthors{Plotkin et al.}

\begin{document}
\def\bl{BL~Lac}
\defcitealias{collinge05}{C05}
\defcitealias{anderson07}{A07}
\defcitealias{plotkin10}{P10}

\newcommand{\nrqsamp}{26}
\newcommand{\noptall}{723}         
\newcommand{\noptrl}{637}           
\newcommand{\noptrq}{86}           
\newcommand{\noptallXray}{294}   

\title{Multiwavelength Observations of Radio-Quiet Quasars With Weak Emission Lines}

\author{
Richard~M.~Plotkin,\altaffilmark{1,2}
Scott~F.~Anderson,\altaffilmark{2} 
W.~N.~Brandt,\altaffilmark{3,4} 
Aleksandar~M.~Diamond-Stanic,\altaffilmark{5} 
Xiaohui~Fan,\altaffilmark{5} 
Chelsea~L.~MacLeod,\altaffilmark{2} 
Donald~P.~Schneider,\altaffilmark{3} 
and Ohad~Shemmer\altaffilmark{6} 
}
\altaffiltext{1}{Astronomical Institute `Anton Pannekoek', University of Amsterdam, Science Park 904, 1098 XH, Amsterdam, the Netherlands; r.m.plotkin@uva.nl}
\altaffiltext{2}{Department of Astronomy, University of Washington, Box 351580, Seattle, WA 98195, USA}
\altaffiltext{3}{Department of Astronomy and Astrophysics, Pennsylvania State University, 525 Davey Laboratory, University Park, PA 16802, USA}
\altaffiltext{4}{Institute for Gravitation and the Cosmos, The Pennsylvania State University, University Park, PA 16802, USA}

\altaffiltext{5}{Steward Observatory, University of Arizona, Tucson, AZ 85721, USA}
\altaffiltext{6}{Department of Physics, University of North Texas, Denton, TX 76203, USA}

\begin{abstract}
We present radio and X-ray observations, as well as optical light curves, for a subset of 26 \bl\ candidates from the Sloan Digital Sky Survey (SDSS) lacking strong radio emission and with $z<2.2$.  Half of these 26 objects are shown to be stars, galaxies, or absorbed quasars.   We conclude that the other 13 objects are Active Galactic Nuclei (AGN) with abnormally weak emission features;  ten of those 13 are definitively radio-quiet, and, for those with available optical light curves, their level of optical flux variability is consistent with radio-quiet quasars.    We cannot exclude the possibility that some of these 13 AGN lie on the extremely radio-faint tail of the \bl\ distribution, but our study generally supports the notion that all \bl\ objects are radio-loud.    These radio-quiet AGN appear to have intrinsically weak or absent broad emission line regions, and, based on their X-ray properties, we argue that some are low-redshift analogs to weak line quasars (WLQs).   SDSS \bl\ searches are so far the only systematic surveys of the SDSS database capable of recovering such exotic low-redshift WLQs.  There are 71 more $z<2.2$ radio-quiet \bl\ candidates already identified in the SDSS not considered here, and many of those might be best unified with WLQs as well.  Future studies combining low- and high-redshift WLQ samples will yield new insight on our understanding of the structure and formation of AGN broad emission line regions.  
\end{abstract}


\section{Introduction}
Strong optical and ultraviolet (UV) emission lines are among the most prominent and often defining characteristics of Active Galactic Nuclei (AGN).  The rare instances of AGN lacking strong emission features are typically classified as BL~Lacertae objects.  In addition to their featureless optical spectra, \bl\ objects are strong radio, X-ray and gamma-ray emitters, they are highly polarized, and they display strong and rapid flux variability across the entire electromagnetic spectrum \citep[e.g., see][]{kollgaard94, perlman01}.    In the standard unification paradigm, \bl\ objects are explained as low-luminosity radio galaxies with a relativistic jet pointed toward the observer \citep[e.g., see][]{blandford78,urry95}.  In this scenario, \bl\ objects' emission lines are diluted by the Doppler boosted relativistic jet, and the \bl\ phenomenon should go hand in hand with strong radio emission.   Observations have indeed established that radio-quiet \bl\ objects must be extraordinarily rare if they exist at all \citep[e.g., see][]{jannuzi93, londish04}.  For example, \citet{stocke90} found no examples of radio-quiet \bl\ objects in the X-ray selected \bl\ sample from the {\it Einstein Observatory} Extended Medium-Sensitivity Survey \citep[EMSS,][]{stocke91}.

It is possible, however, that venerable \bl\ samples such as the EMSS were merely too small ($\sim$40 objects) to reveal especially rare subsets of radio-shy \bl\ objects.   With large-scale multiwavelength surveys and inclusive selection approaches, the number of known \bl\ objects  has grown over the past 10-15 years, from a couple hundred cataloged in \citet{padovani95_mnras} to $\sim$$10^3$ in \citet{massaro09}.  Quasar samples have also become extraordinarily large;  for example, the Sloan Digital Sky Survey \citep[SDSS,][]{york00} Data Release Seven (DR7) Quasar Catalog contains $\sim$10$^5$ quasars with reliable spectroscopic redshifts \citep{schneider10_ph}.   These quasar samples are now large enough to reveal interesting examples of rare quasar populations, such as radio-quiet objects with weak or absent spectral features, in numbers comparable to early \bl\ samples.    For example,  \citet[][hereafter C05]{collinge05} and \citet[][hereafter A07]{anderson07} discovered about two dozen \bl\ candidates lacking strong radio emission.    \citet[][hereafter P10]{plotkin10} present a sample of \noptall\ optically selected \bl\ objects uniformly selected from the SDSS that includes \noptrq\ objects with radio fluxes or limits firmly placing them in the radio-quiet regime.  
 
 This population of weak-featured radio-quiet objects recovered by SDSS \bl\ searches is intriguing, and it is puzzling how nature can create weak-featured AGN that are not strong radio emitters.  If these objects' emission lines are not simply diluted by beamed emission, then some other mechanism must be invoked to explain their apparently anemic broad emission line regions (BELRs).

 Around one-third of the weak-featured radio-quiet AGN recovered by SDSS \bl\ searches  have $z>2.2$. These high-redshift objects pass the phenomenological criteria to alternatively  be classified as weak line quasars (WLQs) - a rare class of high-redshift AGN discovered by the SDSS with weak or absent UV emission features \citep[see][]{fan99,anderson01,shemmer06,shemmer09,diamond09}.     There is significant overlap in the optical properties of  WLQs and \bl\ objects, but WLQs tend to be weaker radio and X-ray emitters and they show less flux variability and polarization.  WLQs appear to compose an exotic population of featureless AGN distinct from \bl\ objects \citep[e.g., see][]{shemmer06,shemmer09,diamond09}.  That is, they seem to have intrinsically weak BELRs, and their emission lines are unlikely to be simply diluted by a beamed relativistic jet.
 
 WLQ redshifts can often be determined only from the onset of the Ly$\alpha$ forest in their SDSS spectra.  For this reason, studies that explicitly search for WLQs in the SDSS database only target high-redshift sources, and $\sim$80 high-redshift WLQs have been discovered by the SDSS to date.    SDSS \bl\ searches are sensitive not only to high-redshift WLQs but also to lower-redshift analogs, if such objects exist.\footnote{\bl\ surveys do not recover all known WLQs because \bl\ searches typically impose more stringent criteria on spectral feature  strengths.}  
 Perhaps some of the lower-redshift weak-featured radio-quiet AGN from SDSS \bl\ searches are actually lower-redshift analogs to high-redshift WLQs.   
  
In this paper we discuss the multiwavelength properties of the population of 26 lower-redshift ($z<2.2$) radio-quiet \bl\ candidates recovered by the \citetalias{collinge05} and \citetalias{anderson07} surveys.  We briefly discuss objects from \citetalias{plotkin10} as well, but we defer detailed description of their properties for another paper. Throughout, we refer to these objects as ``radio-quiet \bl\ candidates'' to reflect the method by which they were discovered, but we will show  that not all objects should be unified with normal radio-loud \bl\ objects.  In Section \ref{sec:ch5_sample} we present the sample of 26 low-redshift radio-quiet \bl\ candidates.    We then present new radio and X-ray observations of a selected subset in Sections~\ref{sec:vla} and \ref{sec:chandra} respectively.  In Section \ref{sec:ch5_var} we discuss their optical variability, and we compare them to radio-loud \bl\ objects as well as to normal radio-quiet quasars.    Contaminants are identified in Section~\ref{sec:notbl}; in Section~\ref{sec:disc} we discuss potential physical interpretations for their weak-featured spectra, we compare them to other lineless AGN in the literature.  Our main conclusions are summarized in Section \ref{sec:conc}.   We adopt the following cosmology: $H_0=71$~km~s$^{-1}$~Mpc$^{-1}$, $\Omega_m=0.27$, and $\Lambda_0=0.73$.

\section{The Parent Radio-Quiet BL~Lac Sample}
\label{sec:ch5_sample}
In Table~\ref{tab:ch5_rquiet} we present all \nrqsamp\ low-redshift ($z<2.2$) radio-quiet \bl\ candidates recovered by \citetalias{collinge05} and \citetalias{anderson07}.    Throughout the text we identify objects that should be removed from this list (also see Section~\ref{sec:notbl}), but we include all \nrqsamp\ objects here for completeness.     Sources with $z>2.2$ are not included in Table~\ref{tab:ch5_rquiet}   because WLQs have been discussed elsewhere \citep[e.g.,][]{shemmer09,diamond09}.    

\tabletypesize{\scriptsize}
\begin{deluxetable*}{c r@{.}l r@{.}l r@{.}l  r@{.}l r@{.}l r@{.}l r@{.}l r@{.}l c l}
\tablewidth{0pt}
\tablecaption{Radio-Quiet \bl\ Candidates \label{tab:ch5_rquiet}}
\tablehead{
	   \colhead{SDSS Name} 
	& \multicolumn{2}{c}{RA} 
	& \multicolumn{2}{c}{Dec} 
	& \multicolumn{2}{c}{$z_{spec}$\tablenotemark{a}} 
	& \multicolumn{2}{c}{$\mu$\tablenotemark{b}} 
	& \multicolumn{2}{c}{$r$\tablenotemark{c}} 
	& \multicolumn{2}{c}{$\log(\nu_oL_o)$\tablenotemark{d}}  
	& \multicolumn{2}{c}{$\alpha_{ro}^{first}$} 
	& \multicolumn{2}{c}{$\alpha_{ox}^{rass}$} 
	& \colhead{Class.\tablenotemark{e}}   
	& Comments\tablenotemark{f}  
\\
	    \colhead{(J2000)} 
	 & \multicolumn{2}{c}{(J2000)}
	 & \multicolumn{2}{c}{(J2000)} 
	 & \multicolumn{2}{c}{} 
	 & \multicolumn{2}{c}{}  
	 & \multicolumn{2}{c}{(mag)} 
	 & \multicolumn{2}{c}{(erg~s$^{-1}$)} 
	 & \multicolumn{2}{c}{} 
	 &  \multicolumn{2}{c}{} 
	 &   
	&    
}	
\startdata
012155.87$-$102037.1  &    20&48281  &   -10&34366  &       0&470  				&     7&6  					&   19&52 	 &                                                44&44     &$<$0&234  &  $>$1&223  						&    U     &   C05,VLA \\
013408.95$+$003102.4  &    23&53731  &     0&51735  &     \multicolumn{2}{c}{\nodata}  &   147&4  				&   19&90  &           \multicolumn{2}{c}{\nodata}    &$<$0&266  &  $>$1&152  						&   S     &    C05,VLA,LC \\
020137.70$+$002535.0  &    30&40709  &     0&42641  &     \multicolumn{2}{c}{\nodata}  &   175&9  				&   19&50  &          \multicolumn{2}{c}{\nodata}     &$<$0&235  &  $>$1&238  						&   S     &     C05,VLA \\
024156.37$+$004351.6  &    40&48490  &     0&73100  &       0&990  				&     3&6  					&   19&59  &                                               45&13     &$<$0&217  &  $>$1&074 					 	&    U    &   C05,VLA,LC \\
024157.36$+$000944.1  &    40&48902  &     0&16226  &      0&790?  				&    54&1  				      	&   20&54  &                                              44&64     &   0&249  &     0&863  							&  RLBL   & C05,A07,VLA,LC \\
025046.47$-$005449.0  &    42&69366  &    -0&91361  &     \multicolumn{2}{c}{\nodata}  &    10&9  				       	&   19&99  &                  \multicolumn{2}{c}{\nodata}  &$<$0&283  &  $>$0&931  						&   S     &    C05,VLA,LC \\
025612.47$-$001057.8  &    44&05197  &    -0&18273  &     \multicolumn{2}{c}{\nodata}  & \multicolumn{2}{c}{\nodata}  	&   20&36  &       \multicolumn{2}{c}{\nodata}     &$<$0&288  &  $>$0&876  						&   U?     &    C05,VLA,LC \\
075551.44$+$352549.8  &   118&96436  &    35&43050  &     \multicolumn{2}{c}{\nodata} &     9&1  					&   19&06  &                            \multicolumn{2}{c}{\nodata}    &$<$0&208  &     1&110  					&   X     & A07\\
090133.42$+$031412.5  &   135&38927  &     3&23681  &       0&459  				&     1&6  					&   18&96  &                                                44&56     &$<$0&203  &  $>$1&224  						&    U    & C05 \\
100847.01$+$114946.9  &   152&19591  &    11&82972  &       0&261  				&    11&3  					&   19&45  &                                               43&83     &$<$0&246  &     0&988  						&    RQBL? &   A07 \\
104833.57$+$620305.0  &   162&13990  &    62&05139  &     \multicolumn{2}{c}{\nodata}  &     5&5  				&   19&85  &                           \multicolumn{2}{c}{\nodata}    &$<$0&281  &  $>$1&330  				&    U?    &   C05 \\
114748.99$+$351350.1  &   176&95416  &    35&23059  &     \multicolumn{2}{c}{\nodata}   &     2&5  				&   20&31  &                            \multicolumn{2}{c}{\nodata} &$<$0&318  &     0&833  					&    S?   &   A07 \\
124225.38$+$642918.9  &   190&60579  &    64&48860  &       0&042  				&     1&8  					&   17&12  &                                               42&86     & \multicolumn{2}{l}{\nodata}  &  $>$1&578  			&    G   &   C05,VLA \\
143139.15$+$600657.9  &   217&91313  &    60&11609  &       0&413  				 &    11&8  				&   20&21  &                                              44&11     &$<$0&277  &     1&246  							&    G    &   A07 \\
150818.96$+$563611.2  &   227&07903  &    56&60314  &      2&052?  				 &     4&5  					&   19&51  &                                               45&16     &$<$0&322  &  $>$1&062  						&    U   &   C05,VLA \\
151115.49$+$563715.4  &   227&81455  &    56&62095  &     \multicolumn{2}{c}{\nodata}  &     6&6  				&   20&05  &                      \multicolumn{2}{c}{\nodata}   &$<$0&285  &  $>$1&378  					&     Abs &   C05 \\
153304.11$+$453526.1  &   233&26716  &    45&59061  &     \multicolumn{2}{c}{\nodata}  &     3&7  				&   20&27  &                \multicolumn{2}{c}{\nodata}    &$<$0&311  &     1&215  						&    RQBL?  &   A07,VLA \\
154515.77$+$003235.2  &   236&31574  &     0&54311  &      1&011?  				 &     5&8  					&   18&81  &                                                45&42     &$<$0&237  &  $>$1&235  						&     U    &  C05,VLA \\
161004.03$+$253647.9  &   242&51681  &    25&61332  &     \multicolumn{2}{c}{\nodata} &     6&5  					&   20&36  &                      \multicolumn{2}{c}{\nodata}    &$<$0&318  &     0&876  					&    X      &   A07,VLA \\
165806.77$+$611858.9  &   254&52822  &    61&31638  & $\ge$1&410? 			 & \multicolumn{2}{c}{\nodata}  &   20&69  &                            $\ge$46&41     &$<$0&025  &  $>$1&747  								&   Abs  & C05\\
211552.88$+$000115.4  &   318&97035  &     0&02097  &     \multicolumn{2}{c}{\nodata}  &    12&6  				&   19&47  &                           \multicolumn{2}{c}{\nodata}    & \multicolumn{2}{c}{\nodata}  &  $>$1&192 	&    WLQ  &   C05,VLA,Chandra,LC \\
212019.13$-$075638.3  &   320&07972  &    -7&94400  &     \multicolumn{2}{c}{\nodata}   &    26&2  				&   19&86  &                           \multicolumn{2}{c}{\nodata}  &$<$0&242  &  $>$1&168  					&    S?   &   C05,VLA \\
213950.32$+$104749.5  &   324&95968  &    10&79711  &       0&296  				&    14&9  					&   20&11  &                                                43&59     & \multicolumn{2}{l}{\nodata}  &     0&985  			&     G   &  C05 \\
224749.57$+$134248.1  &   341&95658  &    13&71338  &      1&175?  				&     6&2  					&   18&27  &                                                45&73     & \multicolumn{2}{l}{\nodata}  &  $>$1&293 		 	&    WLQ &   C05,VLA,Chandra \\
231000.83$-$000516.2  &   347&50346  &    -0&08785  &     \multicolumn{2}{c}{\nodata} &     4&2  					&   19&00  &       \multicolumn{2}{c}{\nodata}     &$<$0&192  &  $>$1&254  						&     U?     &  C05,VLA,LC \\
232428.43$+$144324.3  &   351&11848  &    14&72344  &       1&410  				&     1&2  					&   18&76  &                                               45&65     & \multicolumn{2}{l}{\nodata}  &  $>$0&844  			&     WLQ  &   C05,VLA,Chandra \\

\enddata

\tablenotetext{a}{Tentative redshifts marked with `?'.  All objects have redshift upper limits $z<2.2$, derived by the fact that the Lyman forest is not seen in their SDSS spectra.  If an object lacks a redshift, we assume $z=0.3$ when calculating $\alpha_{ro}^{first}$ and $\alpha_{ox}^{rass}$.}
\tablenotetext{b}{In milli-arcsec yr$^{-1}$, from \citet{munn04}.   We consider proper motion measures to be significantly large if  $\mu>30$~milli-arcsec~yr$^{-1}$  because only 1\% of spectroscopically confirmed SDSS quasars have such large proper motion measures (see Section~\ref{sec:notbl_stars}).}
\tablenotetext{c}{Point spread function magnitude in the SDSS $r$ filter.}
\tablenotetext{d}{At rest frame 5000~\AA.}
\tablenotetext{e}{Our classification (see Sections~\ref{sec:notbl} and \ref{sec:disc}).  Radio-quiet weak-featured AGN are classified as radio-quiet \bl\ objects (RQBL?), low-redshift WLQs (WLQ), or unknown (U).  Contaminants (i.e., objects unlikely to be AGN with intrinsically weak lines) are classified as absorbed AGN (Abs), galaxies (G), radio-loud \bl\ objects (RLBL), stars (S), or they are unlikely the proper optical counterpart to a RASS X-ray source (X).  Question marks denote low-confidence classifications. }
\tablenotetext{f}{C05: listed as a radio-quiet \bl\ candidate in \citet{collinge05}; %
A07: X-ray selected \bl\ candidate lacking radio emission in \citet{anderson07}; %
VLA: we took follow-on 8.4~GHz radio observations with the VLA (see Section~\ref{sec:vla}); %
Chandra: we took follow-on X-ray observations with {\it Chandra} (see Section~\ref{sec:chandra}); %
LC: we have a light curve from SDSS Stripe~82 (see Section~\ref{sec:ch5_var}).%
}
\end{deluxetable*}

We refer the reader to \citetalias{collinge05} and \citetalias{anderson07} for detailed descriptions of their \bl\ selection.  Briefly, both surveys require objects with optical SDSS spectra to not show any emission features with rest-frame equivalent widths ($REW$) stronger than 5~\AA, and all spectra must also show a \ion{Ca}{2}~H/K depression smaller than 40\% \citep[see][]{landt02}.   \citetalias{anderson07} require the additional constraint that all \bl\ candidates must match within 60$''$ to an X-ray source in the {\it ROSAT} All Sky Survey \citep[RASS,][]{voges99,voges00}, with no constraints on their radio emission.    \citetalias{collinge05}  do not explicitly require   radio or X-ray emission for inclusion;  post-selection, their sample is correlated to RASS in the X-ray and to the Faint Images of the Radio Sky at Twenty-cm survey 
(FIRST, \citealt{becker95}) and to the NRAO VLA Sky Survey (NVSS, \citealt{condon98}) in the radio.  Both radio surveys were performed with the Very Large Array\footnote{The National Radio Astronomy Observatory is a facility of the National Science Foundation operated under cooperative agreement by Associated Universities, Inc.}  
 (VLA) at 1.4~GHz.    The 20 radio-quiet \bl\ candidates from \citetalias{collinge05} are not detected in the radio by FIRST/NVSS (or are not in the FIRST footprint), or they have radio detections but are radio-quiet.   Seven objects from \citetalias{anderson07}  lack radio detections in FIRST/NVSS, and one  \citetalias{anderson07}\ object was also identified as a radio-quiet \bl\ candidate by \citetalias{collinge05}. 
 
 The parameters listed in Table~\ref{tab:ch5_rquiet} are taken from the SDSS DR7 database \citep{dr7pap}.  We adopt the spectroscopic redshift from the literature, unless we can identify a better redshift from visual examination of their DR7 spectral reductions.  Of the 26 objects,  13 have spectroscopic redshifts identified from weak spectral features.  Five  of the 13 redshifts are tentative (i.e., they are less certain because their weak emission line identifications are less secure or they show only one emission line that we assume to be \ion{Mg}{2}), which we treat as exact.  One of these tentative redshifts is actually a lower limit derived from possible intervening \ion{Mg}{2} absorption.  All objects in Table~\ref{tab:ch5_rquiet} have redshift upper limits $z<2.2$, or else the Lyman forest would be observed in their SDSS spectra.  We also list morphology and proper motion information in Table~\ref{tab:ch5_rquiet} to help assess if any objects lacking redshifts might be stars.  Proper motions are taken from the SDSS+USNO-B proper motion catalog \citep{munn04}, as listed in the SDSS DR7 database.

Table~\ref{tab:ch5_rquiet} includes broad-band radio to optical ($\alpha_{ro}$) and optical to X-ray ($\alpha_{ox}$) spectral indices for each entry.\footnote{The broad-band spectral index, for $\nu_2>\nu_1$, is defined as $\alpha_{\nu_1\nu_2}=-\log(L_{\nu_2}/L_{\nu_1})/\log(\nu_2/\nu_1)$.  Here, $\alpha_{ro}=-\log(L_o/L_r)/5.08$ and $\alpha_{ox}=-\log(L_x/L_o)/2.60$, where $L_r$, $L_o$, and $L_x$~are the specific luminosities (per unit frequency) at rest-frames 5~GHz, 5000~\AA, and 1~keV respectively \citep{tananbaum79, stocke85}.}   
The $\alpha_{ro}$  values in Table~\ref{tab:ch5_rquiet} are based on FIRST detections (or limits) and SDSS, and the $\alpha_{ox}$ values are derived from RASS and SDSS.  We assume objects with unknown redshifts have $z=0.3$ and redshift lower limits are exact, but we note that $\alpha_{ro}$ and $\alpha_{ox}$ are relatively insensitive to the exact redshift used.   Blank entries for $\alpha_{ro}$ indicate the source is outside the FIRST footprint.   \citetalias{collinge05} and \citetalias{anderson07} use slightly different parameters for estimating broad band spectral indices, so we recalculate them here.   If contamination from the host galaxy to the SDSS optical spectrum is observed, then we decompose the SDSS spectrum into an elliptical galaxy\footnote{\bl\ objects are probably exclusively hosted by elliptical galaxies \citep[e.g., see][]{urry00_hstii}.  Host galaxy contamination can typically be seen if the host galaxy accounts for $\gtrsim$20\% of the observed flux.}  
(using the template of \citealt{mannucci01}) and a power law to measure each object's decomposed AGN optical flux \citepalias[see][]{plotkin10} and its optical spectral index $\alpha_o$ (all local spectral indices are defined as $f_{\nu}\sim \nu^{-\alpha_{\nu}}$).  We use these $\alpha_o$ measures, and we assume radio and X-ray spectral indices of $\alpha_r=-0.27$ and $\alpha_x=1.25$, respectively, to   calculate $\alpha_{ro}$ and $\alpha_{ox}$ at rest-frames 5~GHz, 5000~\AA, and 1~keV.  Limits on radio luminosities are estimated for sources not detected in the radio by FIRST/NVSS by assuming radio flux densities  $f_{1.4GHz} < 0.25 + 5\sigma_{rms}$~mJy, where 0.25 is the CLEAN bias and $\sigma_{rms}$ is the uncertainty of the FIRST survey at the source's location on the sky \citep[see][]{becker95}.  We derive flux density limits from FIRST rather than NVSS because of its higher angular resolution.  The optical magnitude in the SDSS filter closest to rest frame 5000~\AA\ is used for calculating optical luminosities (and we correct for extinction with the \citealt{schlegel98} dust maps). X-ray count rate limits are estimated for X-ray undetected sources as 6 counts divided by the exposure time of  the nearest RASS X-ray source.\footnote{For the undetected RASS sources, we did not find adequate archived pointed X-ray observations (i.e., from {\it SWIFT}, {\it Chandra}, {\it XMM-Newton}, etc.).}  
 We then calculate X-ray luminosities (or limits) at 1~keV using the Portable, Interactive Multi-Mission Simulator \citep[PIMMS,][]{mukai93} correcting for Galactic extinction with the \citet{stark92} \ion{H}{1} maps.   
    
Throughout, we consider objects with $\alpha_{ro}<0.2$ to be radio-quiet for two reasons: 1)  this is consistent with the traditional definition for radio-quiet AGN, that is, radio to optical flux ratios $R<10$ \citep{kellermann89, stocke92}.    2) There are 285 optically selected SDSS \bl\ objects from \citetalias{plotkin10} with both X-ray and radio detections in RASS and FIRST/NVSS; these follow a Gaussian distribution in $\alpha_{ro}$ with $\left<\alpha_{ro}\right>=0.42\pm0.08$.   Thus, any object with $\alpha_{ro}<0.2$ is significantly different from a \bl\ object in its radio properties at the $>$2.75$\sigma$ level ($>$99.7\% one-sided significance).  This implicitly accounts for the non-simultineity of our radio and optical observations, and the fact that we assume the same radio spectral index for every object.   We note this is slightly conservative because SDSS \bl\ objects detected by RASS are among the X-ray brightest and are radio-weaker on average \citep[see][]{plotkin08}.

\section{Multiwavelength Properties}
\subsection{Radio Observations with the VLA}
\label{sec:vla}

We obtained radio observations with the VLA for  17/26  radio-quiet \bl\ candidates from \citetalias{collinge05} and \citetalias{anderson07}.    These objects were targeted because they are among the optically brightest candidates ($g<21.3$), and their FIRST/NVSS radio flux limits are not sensitive enough to place them in the radio-quiet regime (or they do not fall within the FIRST footprint).  Due to scheduling constraints, 5/26 \citetalias{collinge05}/\citetalias{anderson07}  $g<21.3$ radio-quiet \bl\ candidates falling between $07^h < RA < 12^h$ were not observed.  The remaining 4/26 \citetalias{collinge05}/\citetalias{anderson07}  objects  were either too optically faint ($g>21.3$) to efficiently obtain improved $\alpha_{ro}$ constraints with the VLA, or they were already known to have $\alpha_{ro}<0.2$ from their FIRST radio flux limits.  

VLA observations were taken in the D-array configuration on March 30, 2007 for five targets, and in the DnC-array configuration on June 8, 2008 for the other 12.  All observations were performed in the X-band (8.4~GHz) with 26 antennas, with a total of 20-30 minutes integration on each source (using 10 sec integrations).  We used $2\times50$~MHz intermediate frequencies (IFs) centered at $\nu=8435$~MHz and 8485~MHz.   Phase calibrators were observed every 10 minutes, and flux calibrations were tied to either 3C~48 or 3C~286.  A summary of the observations is given in Table~\ref{tab:ch5_vla}.  We observed at 8.4~GHz because many \bl\ objects are brighter at higher radio frequencies (i.e., $\alpha_r\sim-0.27$ typically, \citealt{stickel91}).  This combined with the VLA X-band's superior sensitivity compared to other observing bands makes seeking radio detections for \bl\ candidates most efficient at 8.4~GHz.

\tabletypesize{\scriptsize}
\begin{deluxetable*}{c r@{.}l ccc r@{.}l r@{.}l r@{.}l}
\tablewidth{0pt}
\tablecaption{VLA Observations at 8.4~GHz of Radio-Quiet \bl\ Candidates  \label{tab:ch5_vla}}
\tablehead{
	   \colhead{SDSS Name} 
	& \multicolumn{2}{c}{Redshift} 
	& \colhead{Config\tablenotemark{a}} 
	& \colhead{$\sigma_{rms}$} 
	& \colhead{$f_{vla}$} 
	& \multicolumn{2}{c}{$\alpha_{ro}^{vla}$} 
	& \multicolumn{2}{c}{$\alpha_{ro}^{first}$} 
	& \multicolumn{2}{c}{$\alpha_{ox}^{rass}$}  
\\
	    \colhead{(J2000)} 
	 & \multicolumn{2}{c}{}
	 & \colhead{} 
	 & \colhead{(mJy beam$^{-1}$)} 
	 & \colhead{(mJy)} 
	 & \multicolumn{2}{c}{} 
	 & \multicolumn{2}{c}{} 
	 &  \multicolumn{2}{c}{} 
}	
\startdata
012155.87$-$102037.1\,\,\,               &  0&470                      & DnC  &   0.033  &   \nodata                                   &  $<$0&038  &  $<$0&234                   &  $>$1&223 \\
013408.95$+$003102.4\,\,\,                              & \multicolumn{2}{c}{\nodata} & DnC  &   0.047  &   \nodata    &  $<$0&123  &  $<$0&266                   &  $>$1&152 \\
020137.70$+$002535.0\,\,\,               & \multicolumn{2}{c}{\nodata} & DnC  &   0.032  &   \nodata  		  &  $<$0&058  &  $<$0&235                   &  $>$1&238 \\
024156.37$+$004351.6\,\,\,               &  0&990                      & DnC  &   0.036  &      0.43   				  &     0&109  &  $<$0&217                   &  $>$1&074 \\
024157.36$+$000944.1\,\,\,               &  0&790?                     & DnC  &   0.041  &      2.44    				  &     0&318  &     0&249                   &     0&863 \\
025046.47$-$005449.0\,\,\,  		 & \multicolumn{2}{c}{\nodata} & DnC  &   0.033  &   \nodata   		  &  $<$0&095  &  $<$0&283                   &  $>$0&931 \\
025612.47$-$001057.8\,\,\, 		 & \multicolumn{2}{c}{\nodata}  & DnC  &   0.032  &   \nodata    	 	  &  $<$0&111  &  $<$0&288                   &  $>$0&876 \\
124225.38$+$642918.9\,\,\,                                &  0&042                      & DnC  &   0.035  &   \nodata    		 & $<$-0&063  &   \multicolumn{2}{c}{\nodata}  &  $>$1&578 \\
150818.96$+$563611.2\,\,\,  		 & 2&052?                      & DnC  &   0.032  &   \nodata   			  &  $<$0&128  &  $<$0&322                   &  $>$1&062 \\
153304.11$+$453526.1\,\,\,		               & \multicolumn{2}{c}{\nodata} & DnC  &   0.036  &   \nodata    &  $<$0&133  &  $<$0&311                   &     1&215 \\
154515.77$+$003235.2\,\,\, 		 & 1&011?                      & DnC  &   0.041  &   \nodata    			  &  $<$-0&009  &  $<$0&237                   &  $>$1&235 \\
161004.03$+$253647.9\,\,\,  	         & \multicolumn{2}{c}{\nodata} & DnC  &   0.047  &   \nodata    		  &  $<$0&152  &  $<$0&318                   &     0&876 \\
211552.88$+$000115.4\tablenotemark{b}    & \multicolumn{2}{c}{\nodata} & D    &   0.041  &      0.32          &     0&103  &   \multicolumn{2}{c}{\nodata}  &  $>$1&192 \\
212019.13$-$075638.3\,\,\,  	         & \multicolumn{2}{c}{\nodata} & D    &   0.041  &   \nodata    		  &  $<$0&065  &  $<$0&242                   &  $>$1&168 \\
224749.57$+$134248.1\,\,\,       	 & 1&175?                      & D    &   0.041  &      0.98    			  	  &     0&088  &   \multicolumn{2}{c}{\nodata}  &  $>$1&293 \\
231000.83$-$000516.2\,\,\,  		 & \multicolumn{2}{c}{\nodata}                 & D    &   0.040  &   \nodata    			 &  $<$0&035  &  $<$0&192                   &  $>$1&254 \\
232428.43$+$144324.3\,\,\,		 & 1&410                       & D    &   0.048  &   \nodata    			  	 &  $<$0&011  &   \multicolumn{2}{c}{\nodata}  &  $>$0&844 \\
\enddata
\tablenotetext{a}{Configuration of the VLA during observations.  D-array data were taken on March~30,~2007 as VLA program AP524; DnC-array data were taken on June~8,~2008 as VLA program AP551.}
\tablenotetext{b}{Might be a blend of 2 radio sources, both $\sim$1$''$ from the optical position.}
\end{deluxetable*}

We reduced the data using standard routines in AIPS, flagging anomalous $UV$ data when necessary.  We cleaned the data and created maps with the IMAGR task, using a restoring beam $\sim$8$''$ and $2''\times2''$ pixels.  We estimate background $rms$ noises with TVSTAT, which range from $\sigma_{rms}=0.032 - 0.048$~mJy~beam$^{-1}$.  We require radio detections to have 8.4~GHz flux densities $>5\sigma_{rms}$ within 2$''$ of the SDSS source position.  We detect 4/17 sources in the radio, and we measure their flux densities with TVSTAT using apertures 8-10 pixels on a side.   The largest separation between the radio and SDSS optical positions for these four radio detections is 1.4$''$.

We use the 8.4~GHz flux densities (or limits) combined with SDSS optical fluxes to determine $\alpha_{ro}^{vla}$, which is again calculated between rest-frames 5~GHz and 5000~\AA.  We follow the same methodology as in Section \ref{sec:ch5_sample}, placing 5$\sigma_{rms}$ upper limits on the radio flux densities for the non-detections.  A conservative 5$\sigma_{rms}$ threshold is used for consistency with upper limits placed by the FIRST survey, and because an extraordinary claim of a radio-quiet \bl\ discovery requires very stringent constraints.   We include the values (or limits) of the radio-optical broad-band spectral indices derived from our deeper VLA observations ($\alpha_{ro}^{vla}$) in Table~\ref{tab:ch5_vla}, and we also include the broad-band spectral indices derived from FIRST-SDSS ($\alpha_{ro}^{first}$) and RASS-SDSS ($\alpha_{ox}^{rass}$) for reference.

We find only a single object (SDSS J0241$+$0009) to be radio-loud: it has an 8.4~GHz flux density of 2.44~mJy, and we calculate $\alpha_{ro}^{vla}=0.318$.   This  object also appears in FIRST with a 1.4~GHz flux density of 0.67~mJy (below the nominal FIRST flux limit) and $\alpha_{ro}^{first}$=0.249.  This source did not appear in the FIRST catalog at the times \citetalias{collinge05} and \citetalias{anderson07} were published, which is why we included it in our VLA program.   The discrepancy between $\alpha_{ro}^{first}$ and $\alpha_{ro}^{vla}$ is primarily due to our choice of local radio spectral index $\alpha_{r}=-0.27$.  Combining our 1.4 and 8.4~GHz flux density measurements, we find $\alpha_{r}=-0.7$.  If we adopt $\alpha_r=-0.7$, then $\alpha_{ro}^{first}=0.273$ and $\alpha_{ro}^{vla}=0.277$.    Regardless, SDSS J0241+0009 is definitively radio-loud.  It also has a flat radio spectrum, it is detected in the X-rays by RASS, and we show in Section \ref{sec:ch5_var} that it is variable.    Although it is on the relatively radio-weak tail of the radio-loud \bl\ distribution, its radio-loudness is not extraordinarily small.  It should therefore be considered a normal radio-loud \bl\ object.   

The other three VLA radio detections have $\alpha_{ro}^{vla}$ values that firmly place them in the radio-quiet regime.  We note that one source (SDSS J2115+0001) might actually be a blend of two radio point sources (it is not clear with the spatial resolution of our observations).  Nevertheless, we consider this source as a radio detection, and we report the total combined flux of the (potentially) two components.  Even the combined radio flux confirms SDSS J2115+0001 is radio-quiet.  

We conclude from the radio observations that the majority of weak-featured radio-quiet objects are indeed atypical in their radio properties from the much larger SDSS radio-loud \bl\ population.  Of the 17 objects with deeper VLA coverage, we only classify a single source as a confident radio-loud \bl\ object (SDSS J0241+0009).  
\subsubsection{VLA Observations of High-Redshift WLQs}
Four $z>2.2$ WLQs included in \citetalias{collinge05}'s list of potential radio-quiet \bl\ candidates were included in our VLA programs, and none was detected in the radio.  We list their 8.4~GHz radio properties in Table~\ref{tab:ch5_vla_wlq} for completeness.    Of these four objects, three have deeper X-ray observations in \citet{shemmer09}, and we include those objects' $\alpha_{ox}$ measures derived from their deeper X-ray observations.\footnote{\citet{shemmer09} use slightly different parameters for calculating $\alpha_{ro}$ and $\alpha_{ox}$.  Among the differences, their broad-band spectral indices are defined with opposite sign compared to ours, and they are referenced at 2500~\AA\ in the optical and 2~keV in the X-ray.  They also assume local spectral indices of 0.5 in both the radio and the optical, and they use $\alpha_{\nu}=1.0$ in the X-ray.    We convert their broad-band spectral indices, $\alpha_{ro}^{\prime}$ and $\alpha_{ox}^{\prime}$, to be consistent with our adopted reference frequencies and local spectral indices using approximate transformations $\alpha_{ro}=-1.059\alpha_{ro}^{\prime} + 0.054$ and $\alpha_{ox}=-1.116\alpha_{ox}^{\prime}-0.058$. \label{footnote:sh_aox}}

\tabletypesize{\scriptsize}
\begin{deluxetable*}{c r@{.}l ccc r@{.}l r@{.}l r@{.}l}
\tablewidth{0pt}
\tablecaption{VLA Observations at 8.4~GHz of High-Redshift WLQs \label{tab:ch5_vla_wlq}}
\tablehead{
	   \colhead{SDSS Name} 
	& \multicolumn{2}{c}{Redshift} 
	& \colhead{Config\tablenotemark{a}} 
	& \colhead{$\sigma_{rms}$} 
	& \colhead{$f_{vla}$} 
	& \multicolumn{2}{c}{$\alpha_{ro}^{vla}$} 
	& \multicolumn{2}{c}{$\alpha_{ro}^{first}$} 
	& \multicolumn{2}{c}{$\alpha_{ox}$}  
\\
	    \colhead{(J2000)} 
	 & \multicolumn{2}{c}{}
	 & \colhead{} 
	 & \colhead{(mJy beam$^{-1}$)} 
	 & \colhead{(mJy)} 
	 & \multicolumn{2}{c}{} 
	 & \multicolumn{2}{c}{} 
	 &  \multicolumn{2}{c}{} 
}	
\startdata
031712.23$-$075850.3  			   &  2&699    &    DnC  &   0.035  &   \nodata  & $<$-0&075 &   \multicolumn{2}{c}{\nodata}  &  1&705\tablenotemark{b} \\
121221.56$+$534127.9			   &  3&190    &    DnC  &   0.038  &   \nodata  & $<$-0&108  &  $<$0&072  &  2&095\tablenotemark{b} \\
123743.09$+$630144.9 			   &  3&535    &    DnC  &   0.044  &   \nodata  & $<$-0&113  &  $<$0&058  &  $>$1&671\tablenotemark{c} \\
142505.59$+$035336.2			    & 2&248?  &    DnC  &   0.038  &   \nodata  & $<$-0&058  &  $<$0&126 &  $>$1&236\tablenotemark{d} \\
\enddata
\tablenotetext{a}{Configuration of the VLA during observations.   Data were taken on June~8,~2008 as VLA program AP551.}
\tablenotetext{b}{$\alpha_{ox}$ value taken from {\it Chandra} detection in \citet{shemmer09}, with their reported $\alpha_{ox}$ value converted to conform to our $\alpha_{ox}$ definition (see footnote~\ref{footnote:sh_aox}).}
\tablenotetext{c}{$\alpha_{ox}$ limit taken from {\it XMM-Newton} non-detection in \citet{shemmer09}, with their reported $\alpha_{ox}$ limit converted to conform to our $\alpha_{ox}$ definition (see footnote~\ref{footnote:sh_aox}).}
\tablenotetext{d}{$\alpha_{ox}$ limit taken from RASS non-detection.}
\end{deluxetable*}

\subsection{X-ray Observations with Chandra}
\label{sec:chandra}

Of the 17 radio-quiet \bl\ candidates with deeper VLA observations, 14 only have lower limits on $\alpha_{ox}$ from RASS.   Their RASS limits are not sensitive enough to determine if they are dissimilar to \bl\ objects in the X-ray, and deeper X-ray observations are required to compare their X-ray properties to the larger radio-loud \bl\ population.

We observed three of these 14 objects in the X-ray with {\it Chandra} during Cycle-10.
  The {\it Chandra} targets were chosen as the optically brightest of the five sources with VLA observations taken in 2007 (which were the only VLA observations we had in hand at that time).  {\it Chandra} data were taken with the Advanced CCD Imaging Spectrometer \citep[ACIS;][]{garmire03} at the nominal S3 aimpoint, using faint telemetry mode.  Exposure times of 9.5~ksec, 3.3~ksec, and 5.0~ksec were achieved for SDSS J2115+0001, SDSS J2247+1342, and  SDSS J2324+1443, respectively.   The data were reduced using standard routines in CIAO.\footnote{{\it Chandra} Interactive Analysis of Observations.  See \url{http://cxc.harvard.edu/ciao/.}}  
The {\it Chandra} observations are summarized in Table~\ref{tab:ch5_chandra}.

\tabletypesize{\scriptsize}
\begin{deluxetable*}{c r@{.}l cc  r@{$\times$}l r@{$\times$}l r@{.}l r@{.}l r@{.}l}
\tablewidth{0pt}
\tablecaption{{\it Chandra} Observations  \label{tab:ch5_chandra}}
\tablehead{
	   \colhead{SDSS Name} 
	& \multicolumn{2}{c}{Redshift} 
	& \colhead{Exp. Time} 
	& \colhead{$N_H$\tablenotemark{a}}  
	& \multicolumn{2}{c}{Count Rate\tablenotemark{b}}  
	& \multicolumn{2}{c}{$f_X$\tablenotemark{c}}  
	& \multicolumn{2}{c}{$\alpha_{ox}^{chandra}$} 
	& \multicolumn{2}{c}{$\alpha_{ro}^{vla}$} 
	& \multicolumn{2}{c}{$\alpha_{ox}^{rass}$}  
\\
	    \colhead{(J2000)} 
	& \multicolumn{2}{c}{}  
	 & \colhead{(ksec)} 
	 & \colhead{(cm$^{-2}$)} 
	 & \multicolumn{2}{c}{(counts s$^{-1}$)} 
	 & \multicolumn{2}{c}{(erg s$^{-1}$ cm$^{-2}$)} 
	 & \multicolumn{2}{c}{} 
	 &  \multicolumn{2}{c}{} 
	 & \multicolumn{2}{c}{} 
	 }		
\startdata
211552.88$+$000115.4\tablenotemark{d} &  \multicolumn{2}{c}{\nodata} & 9.5 & $5.83\times10^{20}$ &   $<$5&$10^{-4}$                                 &                   $<$3.2&$10^{-15}$   &   $>$1&905  &   0&103         & $>$1&192 \\
224749.57$+$134248.1\tablenotemark{e} &  1&175?                                   & 3.3  &  $4.58\times10^{20}$ &  $91^{+21}_{-15}$&$10^{-4}$ &     53.4&$10^{-15}$   & 1&552           &   0&088          & $>$1&293 \\
232428.43$+$144324.3\tablenotemark{f}   &  1&410                                     & 5.0  & $4.15\times10^{20}$ &   $12^{+8}_{-4}$\,\,\,&$10^{-4}$ &       7.0&$10^{-15}$   &   1&796         &   $<$0&011   & $>$0&844 \\
\enddata
\tablenotetext{a}{Column density from \citet{stark92}}
\tablenotetext{b}{From 0.5 -- 6.0 keV, assuming a photon power law index $\Gamma=2.25$.}
\tablenotetext{c}{Corrected for Galactic absorption.}
\tablenotetext{d}{Observation taken on Dec 24, 2008; \dataset[ADS/Sa.CXO#obs/10388]{{\it Chandra} ObsID 10388}}
\tablenotetext{e}{Observation taken on Aug 7, 2009; \dataset[ADS/Sa.CXO#obs/10387]{{\it Chandra} ObsID 10387}}
\tablenotetext{f}{Observation taken on May 31, 2009; \dataset[ADS/Sa.CXO#obs/10386]{{\it Chandra} ObsID 10386}}
\end{deluxetable*}

We filter the X-ray observations from 0.5-6.0~keV to reduce the background.  We then measure the number of X-ray counts within 6 pixel circular apertures centered on each source's position with the CIAO routine {\tt dmextract}.  The 6 pixel apertures include $>$90\% of the encircled energy at the S3 aimpoint.\footnote{See the {\it Chandra} Proposers' Observatory Guide; \url{http://cxc.harvard.edu/proposer/POG/}.}  
    The number of background counts are estimated over circular annuli with inner and outer radii of 10 and 20 pixels respectively.   No photons are detected for SDSS J2115+0001; we place an upper limit of 5 photons \citep[99\% confidence interval, see][]{gehrels86} for SDSS J2115+0001, and we estimate a count rate $< 0.0005$~counts~s$^{-1}$.  SDSS J2247+1342 is detected with $30^{+7}_{-5}$ photons ($0.0091^{+0.0021}_{-0.0015}$~counts~s$^{-1}$);  a weak detection is obtained for SDSS J2324+1443, with $6^{+4}_{-2}$ photons ($0.0012^{+0.0008}_{-0.0004}$~counts~s$^{-1}$).\footnote{Quoted errors are 84.13\% one-sided upper and lower confidence limits from \citet{gehrels86}.  These are calculated using Poisson statistics, and they correspond to Gaussian $\pm$1$\sigma$ confidence intervals.}

X-ray fluxes from 0.5-6.0~keV are estimated from the above count rates using PIMMS, correcting for Galactic absorption with the \citet{stark92} \ion{H}{1} maps, and assuming an X-ray spectral index $\alpha_{\nu}=1.25$.  We also estimate X-ray flux densities at 1~keV.  Using the SDSS filter closest to rest-frame 5000~\AA, we find $\alpha_{ox}^{chandra}>1.90$, $\alpha_{ox}^{chandra}=1.55$, and $\alpha_{ox}^{chandra}=1.80$ for SDSS J2115+0001, SDSS J2247+1342, and SDSS J2324+1443, respectively.  

The broad-band spectral indices  derived from {\it Chandra} for these three objects place them on the very X-ray faint tail of the radio-loud \bl\ $\alpha_{ox}$ distribution.  Among the \noptrl\ radio-loud optically selected \bl\ objects in \citetalias{plotkin10}, \noptallXray\ are detected in the X-ray by RASS, with $\left <\alpha_{ox}\right>=1.11\pm0.21$ (note, this implicitly accounts for variability).  The largest $\alpha_{ox}^{rass}$ for an object with a RASS X-ray detection is 1.66;  only two \bl\ objects lacking RASS X-ray detections have larger $\alpha_{ox}^{rass}$ limits.   If these three objects are typical \bl\ objects with abnormally weak radio-emission, then we might have expected their spectral energy distributions to peak at extremely high frequencies (i.e., they would be extreme high-energy peaked \bl\ objects, see \citealt{padovani95_apj}).  However, in this scenario we would expect them to be relatively X-ray bright, which is excluded by our {\it Chandra} observations.  We thus conclude that the objects with deeper {\it Chandra} observations are dissimilar to \bl\ objects in their X-ray properties, but the multiwavelength properties of these three objects are comparable to high-redshift SDSS WLQs (see Section~\ref{sec:disc_wlq}).

Finally, we calculate $\Delta\alpha_{ox}$, the difference between the measured $\alpha_{ox}^{chandra}$ values and that expected from their optical luminosities given the empirical $\alpha_{ox}$--$L_{\nu}$(2500~\AA) relation in Equation~3 of \citet{just07}.  $L_{\nu}$(2500~\AA) is the specific luminosity at 2500~\AA\ rest-frame, and for our $\Delta \alpha_{ox}$ measurements we re-calculate $\alpha_{ox}^{chandra}$ at rest-frames 2~keV and 2500~\AA\ for consistency with  \citet{just07}.  We find $\Delta \alpha_{ox}$=-0.052 (-0.26$\sigma$) and 0.197 (0.994$\sigma$) for SDSS 2247+1342 and SDSS J2324+1443, respectively. The values in parentheses are normalized to the standard deviation in $\alpha_{ox}$, $\sigma=0.198$, for AGN with optical luminosity densities between 10$^{30}$--10$^{31}$~erg~s$^{-1}$~Hz$^{-1}$ \citep[see Table~5 of ][]{steffen06}.  The lack of a spectroscopic redshift (or an X-ray detection) for SDSS J2115+0001 precludes us from estimating its $\Delta \alpha_{ox}$ value.  The measured $\Delta \alpha_{ox}$ for SDSS 2247+1342 and SDSS J2324+1443 are not significantly different from zero, and we thus conclude that their X-ray properties are similar to normal radio-quiet quasars and WLQs \citep[see][]{shemmer09}.

\subsection{Optical Flux Variability}
\label{sec:ch5_var}
\begin{figure*}
\centering
\begin{minipage}[t]{0.45\linewidth}
\includegraphics[scale=0.43]{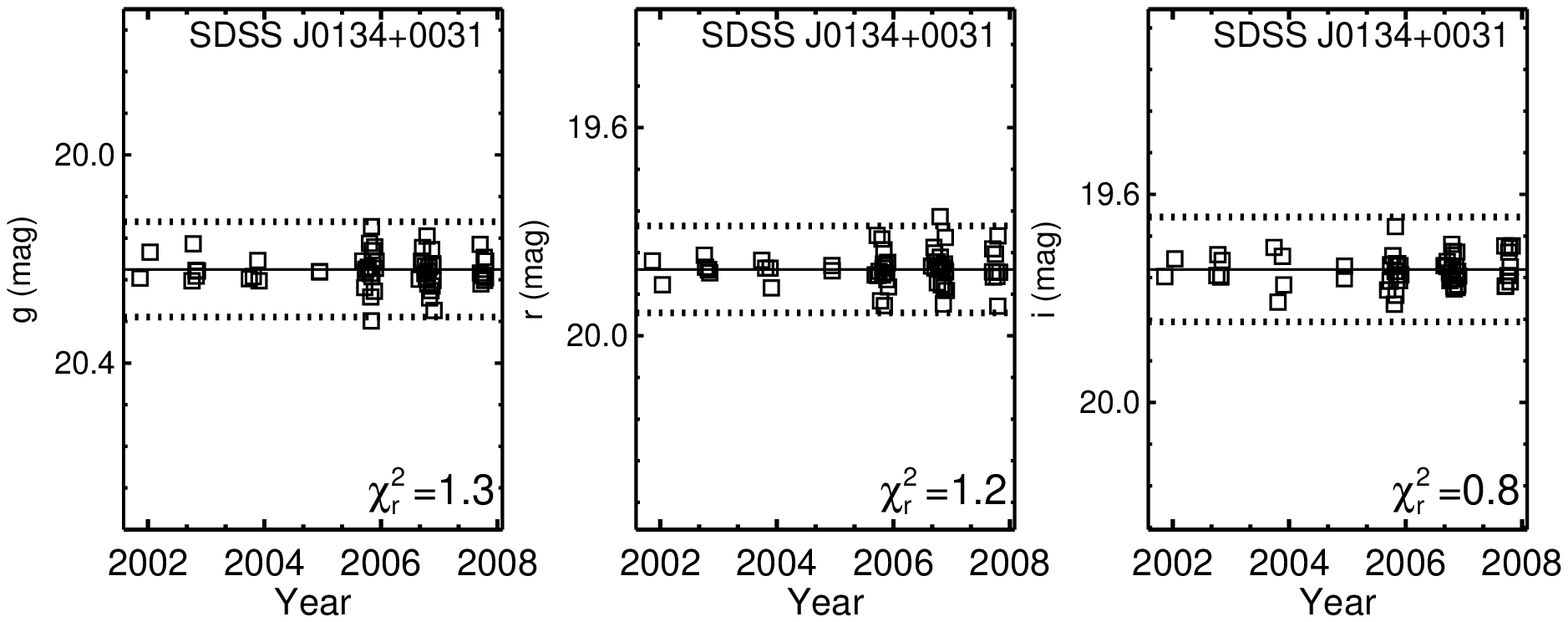}
\includegraphics[scale=0.43]{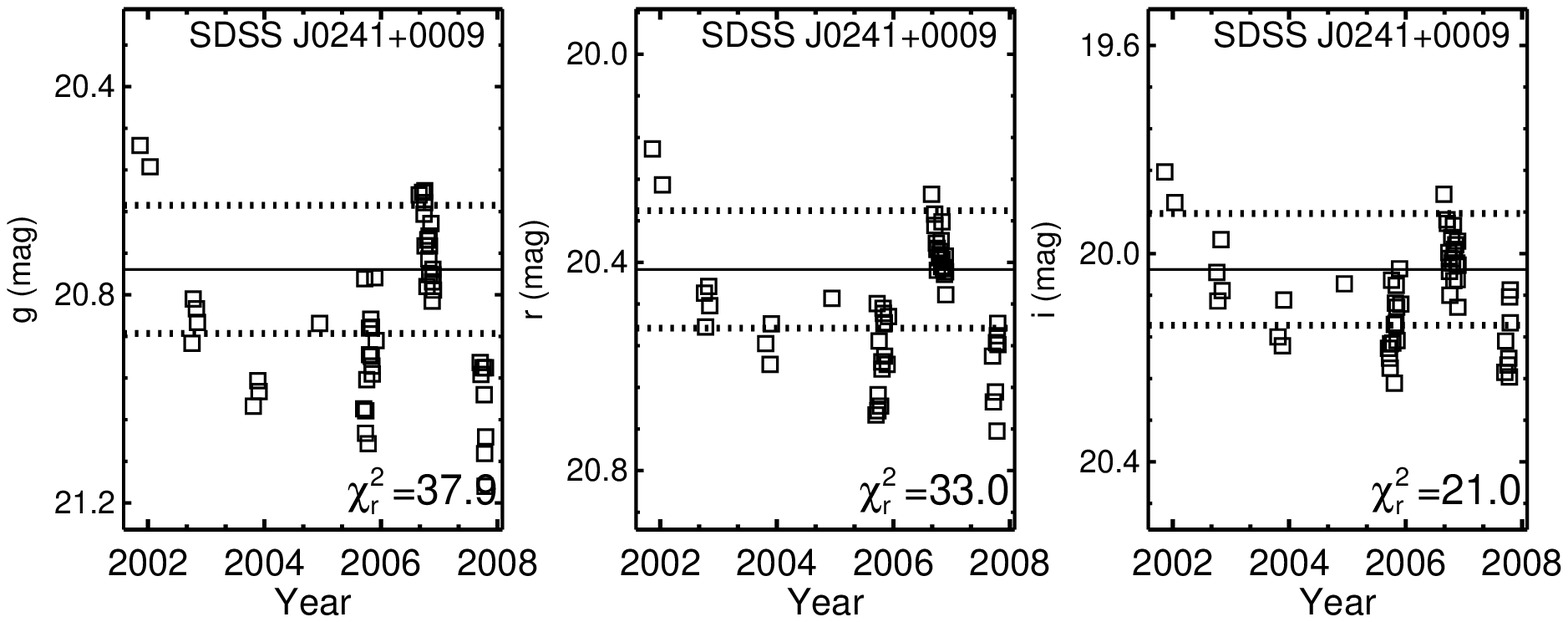}
\includegraphics[scale=0.43]{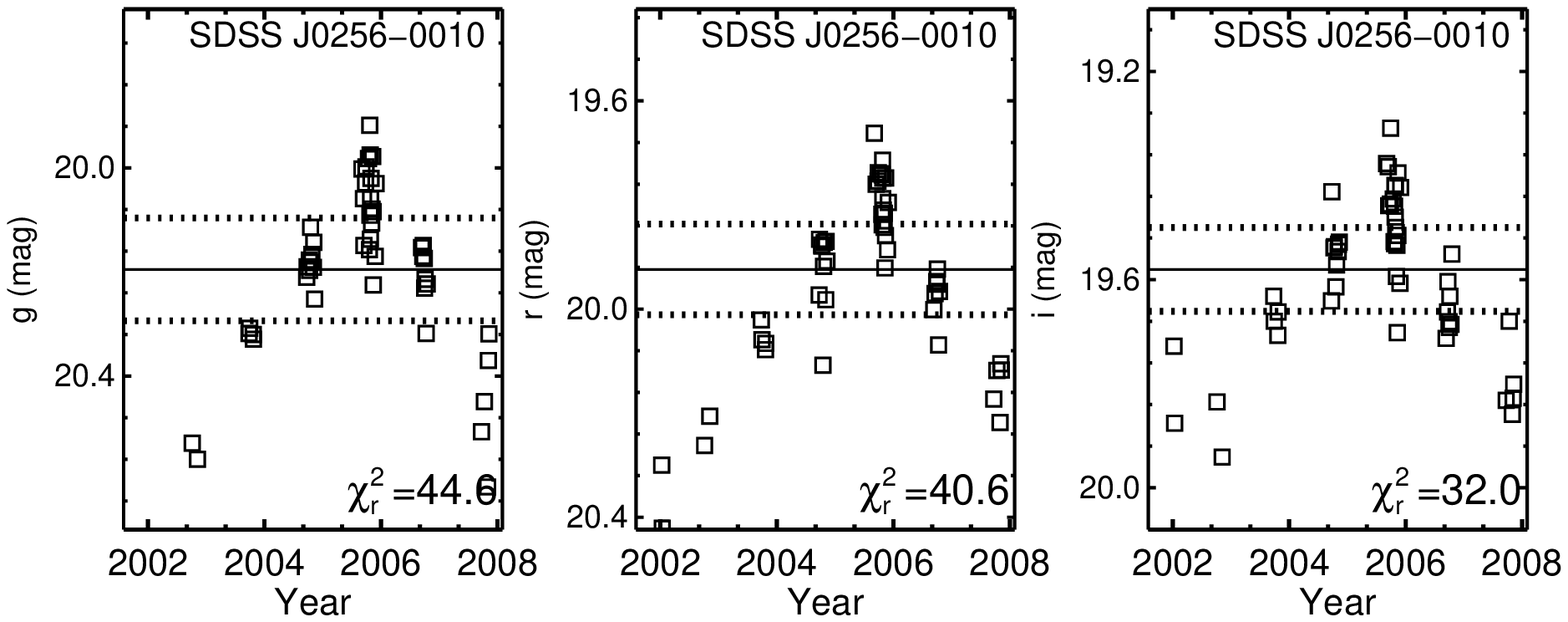}
\includegraphics[scale=0.43]{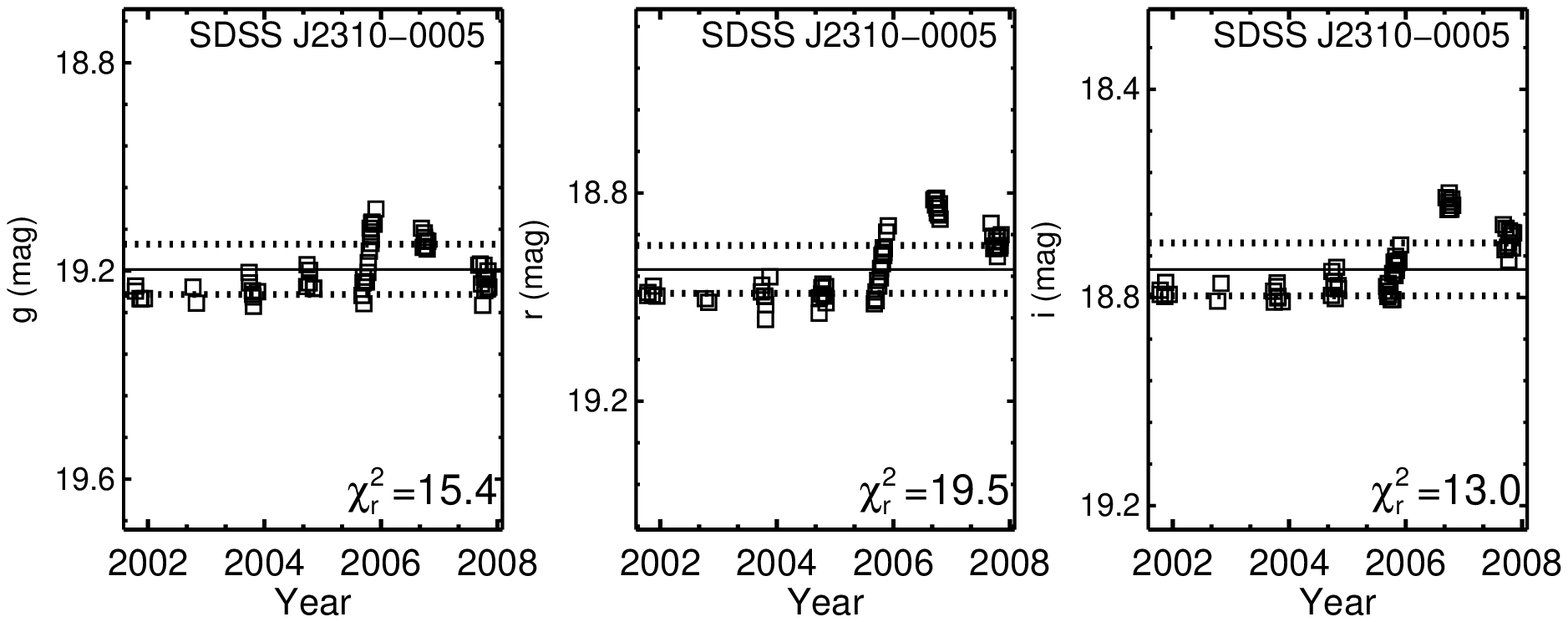}
\end{minipage}
\hspace{0.55cm}
\begin{minipage}[t]{0.45\linewidth}
\includegraphics[scale=0.43]{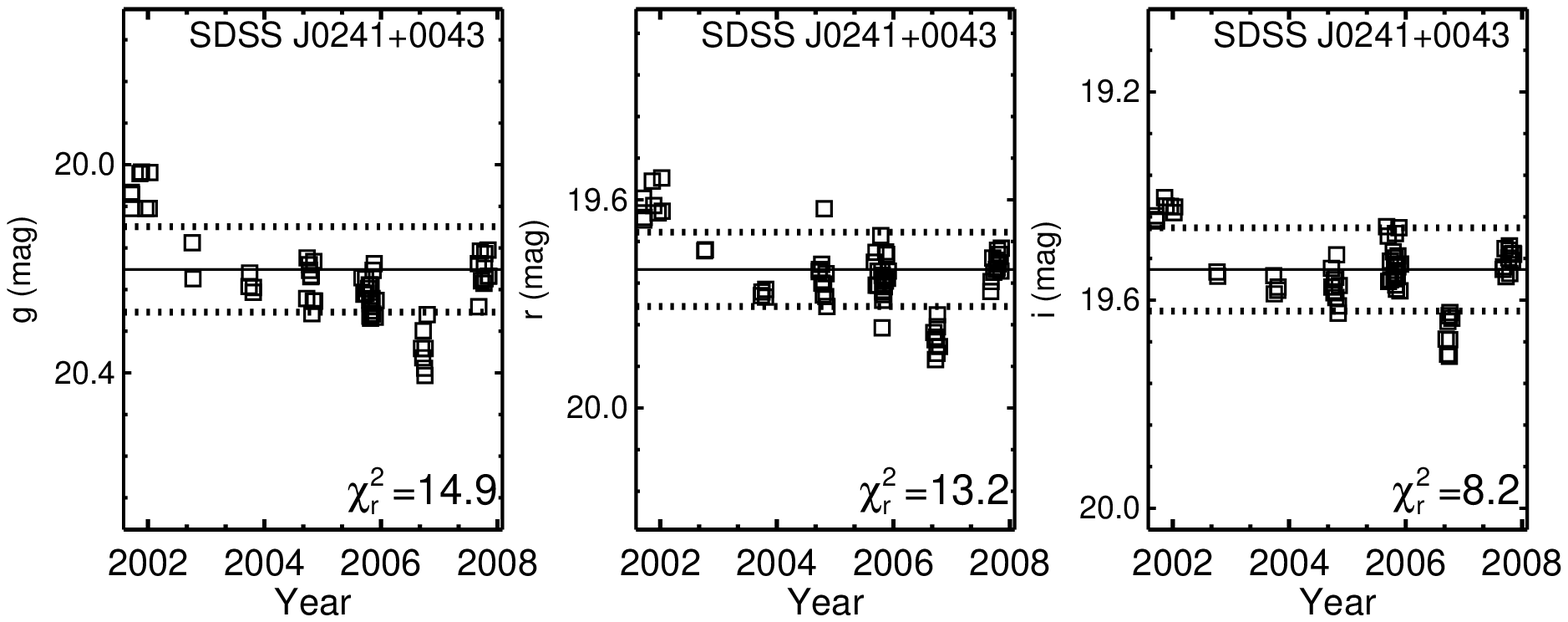}
\includegraphics[scale=0.43]{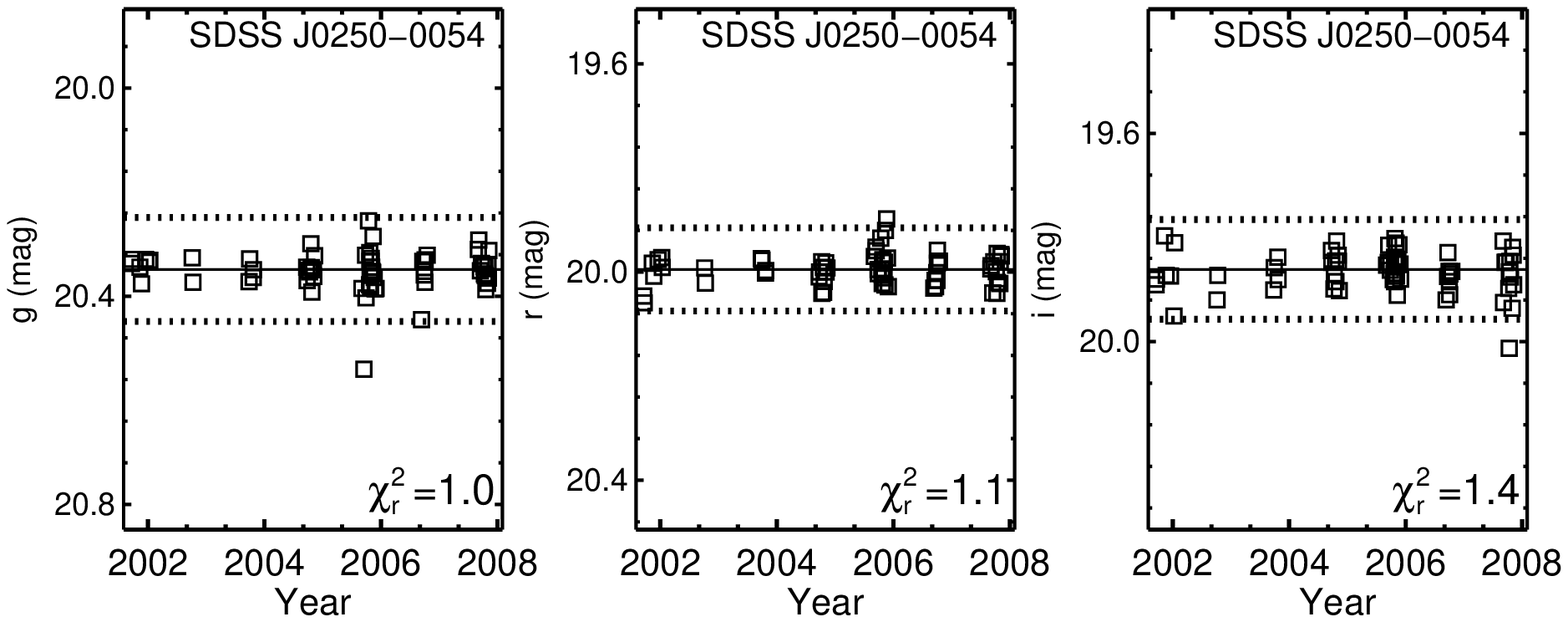}
\includegraphics[scale=0.43]{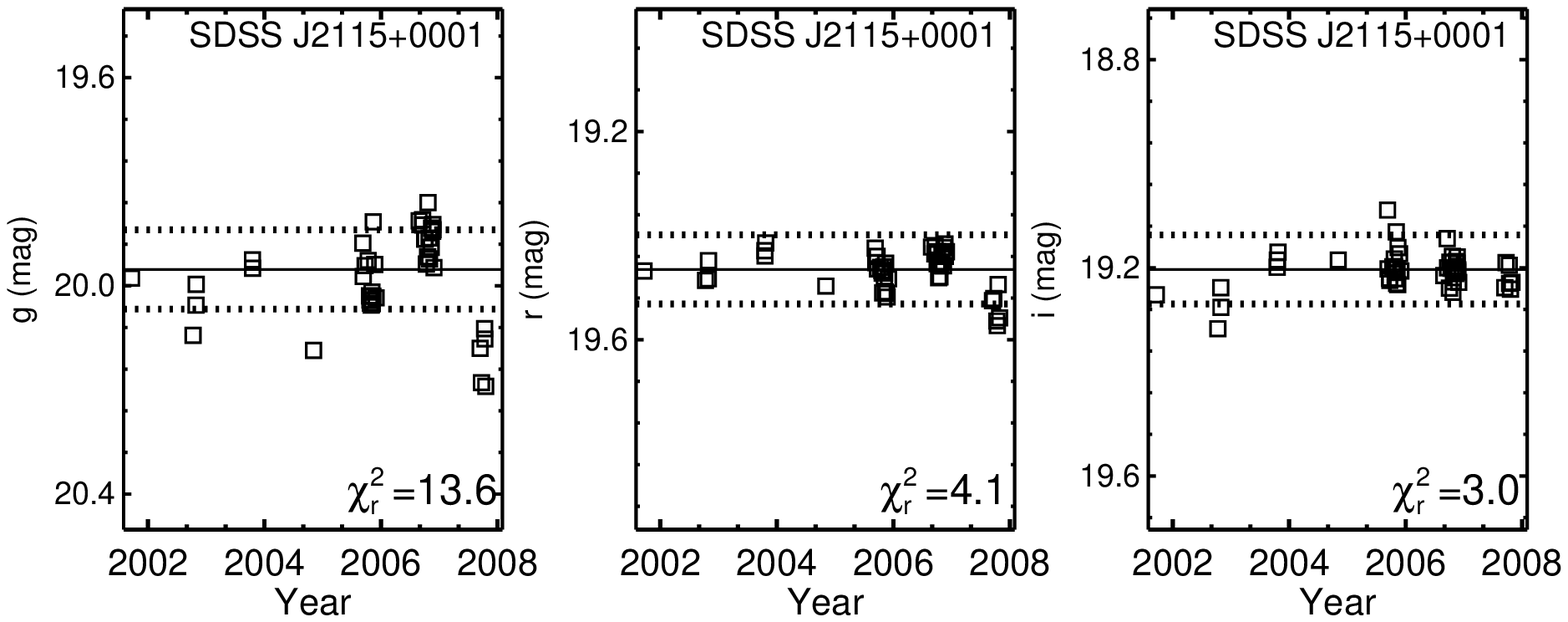}
\end{minipage}

\caption{Stripe~82 light curves in the $g$, $r$, and $i$ filters for seven radio-quiet \bl\ candidates; data taken prior to Fall~2001 are not shown.   The solid lines illustrate the average magnitude of each source, and the dotted lines mark $\pm$3 times the average photometric error.  All panels span 1~mag along the y-axis.   The two non-variable objects (with $\chi_r^2<3$)  are likely stars.}
\label{fig:ch5_lc}
\end{figure*}

Finally, we consider flux variability, as we expect strong variability if these are in fact \bl\ objects.    The SDSS has  observed a $\sim$300~deg$^2$ region of the sky  ($20^{\rm h} 34^{\rm m} < \alpha_{\rm J2000.0} < 04^{\rm h}00^{\rm m}$, $\left |\delta_{\rm J2000.0} \right | < 1.266^{\circ}$) in all five SDSS filters multiple times every Fall since 1998.   This region is referred to as Stripe~82 \citep[see][]{ivezic07,sesar07}.   Seven of our low-redshift radio-quiet \bl\ candidates lie inside Stripe~82, each one with 50 -- 70 epochs of observations (see Figure~\ref{fig:ch5_lc}).   The $u$ and $z$ filters tend to be noisier, so we do not consider them here.    Most objects only have 1-2 epochs of observations per year prior to Fall~2001, so we do not show these data in Figure~\ref{fig:ch5_lc}, but we do include pre-2001 epochs in the following analysis.  

We use the reduced $\chi_r^2$ statistic to assess the level of flux variability in a given filter.\footnote{$\chi_r^2 = \sum_i^N \left[ \left (m_i - \left<m\right>\right )/\sigma_i\right]^2/(N-1)$, where $m_i$ and $\sigma_i$ are the observed magnitudes and uncertainties, respectively, in a given filter, $\left <m \right>$ is the average magnitude, and $N$ is the number of epochs in that filter.}  
If an object has $\chi_r^2>3$, then we consider it to show variability \citep[e.g., see][]{sesar07}.  We find five of the seven sources to be variable in the $g$ filter (SDSS J0241$+$0043, SDSS J0241$+$0009, SDSS J0256$-$0010, SDSS J2115$+$0001, and SDSS J2310$-$0005).  These five objects show flux variations over short and long timescales (ranging from days to years),  and flux variations tend to be correlated between filters.   We note that one of the variable objects is SDSS J0241$+$0009, which we classify as a radio-loud \bl\ object based on its radio and X-ray properties.

The measured values of $\chi_r^2$ and the $rms$ scatter, $\sigma_{rms}$, for each source in each filter are listed in Table~\ref{tab:ch5_lc}.  We include $\alpha_{ro}^{vla}$ and $\alpha_{ox}^{rass}$ for reference, and we list the average measured photometric error for each source in each filter, $\left <\sigma_m\right>$.  Observations typically have photometric uncertainties of $\sim$0.03~mag, and the largest degree of variability we detect is $\sigma_{rms}\sim0.2$~mag (SDSS J0256$-$0010).   The two objects that do not appear variable are likely stars.   Neither has a spectroscopic redshift, and one of them (SDSS J0134+0031) has a very large proper motion measure ($\mu\sim 150$~milli-arcsec~yr$^{-1}$).

\tabletypesize{\small}
\begin{deluxetable*}{c r@{.}l r@{.}l r@{.}l c r@{.}l cc}
\tablewidth{0pt}
\tablecaption{Optical Variability of Seven Objects in Stripe 82 \label{tab:ch5_lc}}
\tablehead{
	    \colhead{SDSS Name} 
	 & \multicolumn{2}{c}{Redshift} 
	& \multicolumn{2}{c}{$\alpha_{ro}^{vla}$} 
	& \multicolumn{2}{c}{$\alpha_{ox}^{rass}$}   
	& \colhead{filter}   
	& \multicolumn{2}{c}{$\chi_r^2$}               
	& \colhead{$\sigma_{rms}$}   
	& \colhead{$\left < \sigma_{m}\right >$}       
\\
	    \colhead{(J2000)} 
	  & \multicolumn{2}{c}{} 
	 & \multicolumn{2}{c}{}
	 & \multicolumn{2}{c}{} 
	 & \colhead{} 
	 & \multicolumn{2}{c}{} 
	 & \colhead{(mag)} 
	 &  \colhead{(mag)}
}	
\startdata

013408.95$+$003102.4  &    \multicolumn{2}{c}{\nodata} & $<$0&123  			&    $>$1&152  				      &     $g$  &     1&293  &     0.033  &     0.031 \\
             \nodata  &   \multicolumn{2}{c}{\nodata}  &       \multicolumn{2}{c}{\nodata}  &            \multicolumn{2}{c}{\nodata}  &     $r$  &      1&221  &     0.032  &     0.028 \\
             \nodata  &   \multicolumn{2}{c}{\nodata}  &       \multicolumn{2}{c}{\nodata}  &            \multicolumn{2}{c}{\nodata}  &     $i$  &      0&777  &     0.029  &     0.034 \\ \hline
024156.37$+$004351.6  &    0&990                                    &  0&109				&    $>$1&074  				     &     $g$  &    14&853  &     0.085  &     0.027 \\
             \nodata  &   \multicolumn{2}{c}{\nodata}  &       \multicolumn{2}{c}{\nodata}  &     \multicolumn{2}{c}{\nodata}        &     $r$  &    13&192  &     0.074  &     0.024 \\
             \nodata  &   \multicolumn{2}{c}{\nodata}  &       \multicolumn{2}{c}{\nodata}  &     \multicolumn{2}{c}{\nodata}        &     $i$  &       8&167  &     0.066  &     0.027 \\ \hline
024157.36$+$000944.1  &     0&790?                    &           0&318  		                &       0&863  				     &     $g$  &    37&908  &     0.185  &     0.041 \\
             \nodata  &   \multicolumn{2}{c}{\nodata}  &       \multicolumn{2}{c}{\nodata}  &     \multicolumn{2}{c}{\nodata}        &     $r$  &    32&990  &     0.162  &     0.038 \\
             \nodata  &  \multicolumn{2}{c}{\nodata}  &        \multicolumn{2}{c}{\nodata}  &     \multicolumn{2}{c}{\nodata}        &     $i$  &    20&972  &     0.130  &     0.036 \\ \hline
025046.47$-$005449.0  &    \multicolumn{2}{c}{\nodata} &$<$0&095 			&    $>$0&931				     &     $g$  &     0&973  &     0.038  &     0.033 \\
             \nodata  &  \multicolumn{2}{c}{\nodata}  &        \multicolumn{2}{c}{\nodata}  &           \multicolumn{2}{c}{\nodata}  &     $r$  &     1&086  &     0.030  &     0.027 \\
             \nodata  &  \multicolumn{2}{c}{\nodata}  &        \multicolumn{2}{c}{\nodata}  &           \multicolumn{2}{c}{\nodata}  &     $i$  &     1&398  &     0.040  &     0.032 \\ \hline
025612.47$-$001057.8  &    \multicolumn{2}{c}{\nodata}  &       $<$0&111			     &    $>$0&876   		    &     $g$  &    44&585  &     0.203  &     0.033 \\
             \nodata  &  \multicolumn{2}{c}{\nodata}  &        \multicolumn{2}{c}{\nodata}       &     \multicolumn{2}{c}{\nodata}  &     $r$  &    40&551  &     0.171  &     0.029 \\
             \nodata  &  \multicolumn{2}{c}{\nodata}  &        \multicolumn{2}{c}{\nodata}       &     \multicolumn{2}{c}{\nodata}  &     $i$  &    31&963  &     0.147  &     0.027 \\ \hline
211552.88$+$000115.4  &   \multicolumn{2}{c}{\nodata}  & 0&103  &    $>$1&905\tablenotemark{a}  &     $g$  &    13&577  &     0.085  &     0.025 \\
             \nodata  &  \multicolumn{2}{c}{\nodata}  &        \multicolumn{2}{c}{\nodata}  &     \multicolumn{2}{c}{\nodata}    &     $r$  &     4&080  &     0.040  &     0.022 \\
             \nodata  &  \multicolumn{2}{c}{\nodata}  &        \multicolumn{2}{c}{\nodata}  &     \multicolumn{2}{c}{\nodata}    &     $i$  &     2&988  &     0.037  &     0.022 \\ \hline
231000.83$-$000516.2  &   \multicolumn{2}{c}{\nodata}          &         $<$0&035  &    $>$1&254  &     $g$  &    15&379  &     0.050  &     0.016 \\
             \nodata  &  \multicolumn{2}{c}{\nodata}  &        \multicolumn{2}{c}{\nodata}  &     \multicolumn{2}{c}{\nodata}   &     $r$  &    19&540  &     0.066  &     0.015 \\
             \nodata  &  \multicolumn{2}{c}{\nodata}  &        \multicolumn{2}{c}{\nodata}  &     \multicolumn{2}{c}{\nodata}   &     $i$  &    12&967  &     0.062  &     0.017 \\ 
\enddata
\tablenotetext{a}{X-ray flux limit from {\it Chandra}}
\end{deluxetable*}

The detection of variability does not necessarily indicate that these are \bl\ objects, because radio-quiet quasars are also variable.  We now compare the level of variability of the radio-quiet \bl\ candidates to normal radio-loud \bl\ objects and to radio-quiet quasars.  There are 14 optically-selected \bl\ candidates in \citetalias{plotkin10} with Stripe~82 light curves that are detected in the radio by FIRST/NVSS and are radio-loud; we calculate $\chi_r^2$ for those objects in the $g$-filter.\footnote{Because SDSS J0241+0009 has radio-loud $\alpha_{ro}$ measures from both FIRST and our follow-up VLA observations, we do not consider it a radio-quiet \bl\ candidate in the following.  It was recovered as a radio-loud \bl\ candidate in \citetalias{plotkin10}, and it is thus included in the 14 object subset of radio-loud \bl\ objects with Stripe~82 coverage.}  
     We also examine a sample of $\sim$2000 spectroscopically confirmed radio-quiet SDSS DR5 quasars in Stripe~82 from \citet{schneider07}, and we measure $\chi_r^2$ for each of these radio-quiet quasars.   These $\sim$2000 quasars are selected because: 1) they have Stripe~82 light curves with at least 10 epochs of data in each filter; 2) they are in the FIRST survey's footprint and have $\alpha_{ro}<0.2$, derived from FIRST radio detections or limits; (3) they have $g>18.7$, so that their optical fluxes are similar to our radio-quiet \bl\ candidates with Stripe~82 coverge; and (4) they have $0.5<z<2.2$.  This redshift range encompasses the redshifts of our seven radio-quiet \bl\ candidates in Stripe~82.   All have $z<2.2$, since we do not see the Ly$\alpha$ forest in their SDSS spectra; those lacking spectroscopic redshifts probably have $z>0.5$, or else we would likely see host galaxy spectral features in their optical spectra (see \citetalias{plotkin10}).    We restrict our comparison quasar sample to these redshifts so that our $\chi^2_r$ measures are calculated at approximately similar rest-frames.    For simplicity, the following analysis is illustrated with values derived in the $g$ filter, but our conclusions do not change if we use a different filter (or if we choose filters to cover similar rest-frame wavelengths).

\begin{figure}
\centering
\includegraphics[width=90mm,height=75mm]{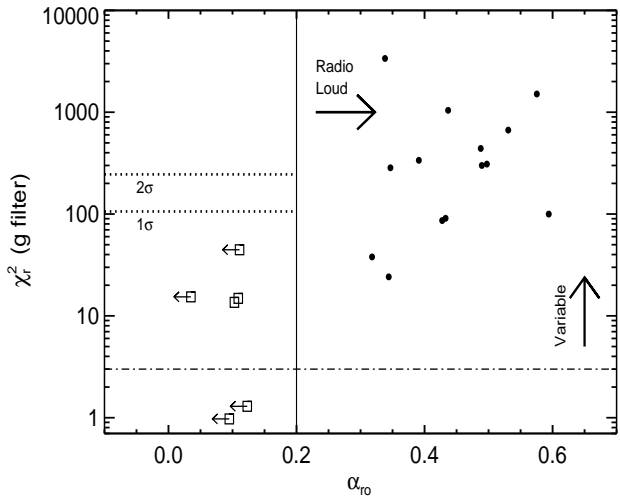}  
\caption{Reduced $\chi_r^2$ of flux variability in the $g$ filter vs.\ $\alpha_{ro}$.   {\it Circles:} 14 radio-loud \bl\ objects from \citetalias{plotkin10} with Stripe~82 light curves.  {\it Squares:} six radio-quiet \bl\ candidates with Stripe~82 light curves and $\alpha_{ro}$ measures (or limits) smaller than 0.2.  The vertical solid line shows the traditional boundary between radio-loud and radio-quiet AGN.  The horizontal dashed-dotted line shows $\chi_r^2=3$, above which we conclude a source is variable.    The dotted lines mark the maximum values of $\chi_r^2$ we expect 84\% (1$\sigma$) and 97.5\% (2$\sigma$) of radio-quiet quasars to show.  The variability of the radio-quiet \bl\ candidates is consistent with that expected for radio-quiet quasars, although there is some overlap between the high  and low variability tails of the radio-quiet quasar and \bl\ distributions.   This figure is similar if we rather plotted $\chi_r^2$ for the  $r$ or $i$ filters.}
\label{fig:ch5_dr7chisq}
\end{figure}

Figure~\ref{fig:ch5_dr7chisq} shows $\chi_r^2$ vs.\ $\alpha_{ro}$ for the  six radio-quiet \bl\ candidates with Stripe~82 light curves (squares, not including SDSS J0241+0009) and for the 14  radio-loud \bl\ candidates (circles) from \citetalias{plotkin10} that lie in Stripe~82.  We use VLA data when available to calculate $\alpha_{ro}$.    The $\sim$2000 radio-quiet quasars have an average $\left<\chi_r^2\right>=65$,  and 84\% (i.e., the one-sided 1$\sigma$ confidence interval) have  $\chi_r^2<106$, and 97.5\% (i.e., the one-sided 2$\sigma$ confidence interval) have $\chi_r^2<245$ (marked by the dotted lines in Figure~\ref{fig:ch5_dr7chisq}).\footnote{Only 19 quasars have $\chi_r^2<3$  in the $g$ filter.}  
There is a marked deficit of extremely variable radio-quiet objects.

The level of flux variability is likely dominated by different physical processes in radio-quiet and radio-loud quasars.     Using a sample of $\sim$100 quasars covering a wide range of redshifts, optical luminosities, and radio loudnesses, \citet{kelly09} suggest the flux variability of radio-quiet quasars is caused by a stochastic process; this conclusion is confirmed by \citet{macleod10_ph}.  One potential mechanism proposed by \citet{kelly09} to explain the observed variability for radio-quiet quasars is turbulent magnetic fields within the accretion disk.    For radio-loud quasars, however, additional mechanisms related to the radio jet need to be invoked.  Here, variability is commonly modeled as shocks propagating through an inhomogeneous jet \citep[e.g., see][]{marscher85, ulrich97, aller99}.  

The $\chi_r^2$ values in Figure~\ref{fig:ch5_dr7chisq} suggest that the mechanism driving variability in the radio-quiet \bl\ candidates is similar to that for radio-quiet quasars.       To test this assertion, we assume the observed distribution of $\chi_r^2$  for the four variable radio-quiet \bl\ candidates is randomly drawn from the $\chi_r^2$ distribution of the $\sim$2000 DR5 radio-quiet quasars.  We ignore the two non-variable radio-quiet \bl\ candidates because they are likely stellar.  We also assume the 14 radio-loud \bl\ objects'  $\chi_r^2$ measurements are randomly drawn from the DR5 quasars.  The above null hypotheses are then tested via Monte Carlo simulations, as described below.   This test is designed to achieve the same purpose as a Kolmogorov-Smirnov test, but we use Monte Carlo simulations because of the small sample sizes. 

 Of the 14 \citetalias{plotkin10} radio-loud \bl\ objects with Stripe~82 light curves,  12/14 have $\chi_r^2>65$ (the mean of the DR5 quasars), 9/14 have $\chi_r^2>106$ (1$\sigma$), and 9/14 have $\chi_r^2>245$ (2$\sigma$).  To test the likelihood that the $\sim$2000 DR5 quasars are the parent population of the 14  radio-loud \bl\ objects,  we randomly choose 14 $\chi_r^2$ values from the DR5 quasar distribution.  We then count the number of randomly selected $\chi_r^2$ values larger than 65, 106, and 245; we repeat 10$^6$ times.   Our Monte Carlo simulations show there are p=0.000091, 0.000061, and $<$0.000001 chances of randomly drawing 12, 9, and 9 objects with $\chi_r^2$ values larger than 65, 106, and 245, respectively.  We thus conclude the radio-loud \bl\ objects are statistically different from the DR5 quasars in their variability properties.
 
For the four radio-quiet \bl\ candidates, no object has measured $\chi_r^2>65$.   We perform a similar test, this time randomly drawing only four $\chi_r^2$ values from the distribution of DR5 quasars, and we again repeat 10$^6$ times.  We find there are p=0.19, 0.50, and 0.90 chances of randomly choosing zero quasars with $\chi_r^2>$65, 106, and 245, respectively.  Thus, we conclude it is plausible the level of variability displayed by our radio-quiet \bl\ candidates is similar to that of normal radio-quiet quasars.

We of course caution that one should not draw strong statistical conclusions from these small sample sizes;  however, our Monte Carlo simulations suggest the radio-quiet \bl\ candidates are similar to radio-quiet quasars, but the radio-loud \bl\ objects are not.  This conclusion is strengthened after also considering their radio and X-ray properties.  It is also important to mention that there is overlap in $\chi_r^2$ between the most variable radio-quiet quasars and the least variable radio-loud \bl\ objects.  Thus, we cannot exclude that some of the more variable radio-quiet \bl\ candidates (e.g., SDSS J0256-0010) have weakly boosted continua, but that does not appear to be the dominant source of radiation.  A similar conclusion was reached for some relatively radio-bright high-redshift WLQs \citep[e.g., SDSS J1408+0205 and SDSS J1442+0110;][]{diamond09}.   The most variable radio-quiet \bl\ candidates  should be placed at high priority for polarimetric follow-up.  

\section{Recognition of Sources Unlikely to be Lineless AGN}  
\label{sec:notbl}
Equipped with the above observations, we now identify objects that are unlikely to be AGN with intrinsically weak emission lines.   Here we discuss all 26 radio-quiet \bl\ candidates in Table~\ref{tab:ch5_rquiet}, not just the 17 with deeper VLA observations.  Improvements  to the SDSS data reduction pipelines are implemented prior to each  data release, so we also inspected the DR7 spectra of the  \citetalias{collinge05} and \citetalias{anderson07} radio-quiet \bl\ candidates (which were selected from earlier SDSS data releases). 

\subsection{Normal Radio-Loud \bl\ Object}
\label{sec:notbl_rlbl}
SDSS J0241+0009 is a normal radio loud \bl\ object based on its radio and X-ray properties, and its optical variability.  

\subsection{Stars}
\label{sec:notbl_stars}
   Three of the 26 radio-quiet \bl\ candidates  show large proper motion in Table~\ref{tab:ch5_rquiet}, which we define as $\mu > 30$~milli-arcsec~yr$^{-1}$ \citep[less than 1\% of spectroscopically confirmed SDSS quasars have proper motion measures larger than 30~milli-arcsec~yr$^{-1}$; ][]{schneider07}.  One of these measurements is spurious: SDSS J0241+0009 ($\mu=54$~milli-arcsec~yr$^{-1}$) is a very convincing radio-loud \bl\ object (see Section~\ref{sec:notbl_rlbl}).  The other two sources with large proper motion measures (SDSS J0134+0031 and SDSS J0201+0025) both have $\mu > 100$~milli-arcsec~yr$^{-1}$, they lack spectroscopic redshifts, and they have point source optical morphologies.\footnote{Both objects are from \citetalias{collinge05}, and their proper motions were less certain at the time of publication.}  
 Neither is detected by our deeper radio observations or by RASS in the X-ray.   We have an optical light curve for SDSS J0134+0031 (see Section \ref{sec:ch5_var}), and it does not show variability over an entire decade.  Those two proper motion measures are likely not spurious, and those two objects are removed from the sample as potential stars.   We also remove SDSS J0250-0054 as a likely star because it does not show flux variability in its Stripe~82 light curve, it does not have a measured redshift, and it is not detected in our 8.4~GHz VLA observations.  The spectrum of SDSS J1147+3513 might be contaminated by a bright nearby star, and SDSS J2120$-$0756 might also show weak stellar absorption features in its DR7 spectrum.   We thus remove those two sources as well (neither has a measured redshift).  In all, five of the 26 radio-quiet \bl\ candidates are removed because they are probably stars.

\subsection{Galaxies}
Examination of the improved SDSS spectral reductions of SDSS J1431+6006 and SDSS J2139+1047 show they have \ion{Ca}{2}~H/K breaks larger than 40\%, and we re-classify them as elliptical galaxies.  We also remove SDSS J1242+6429, which is probably a galaxy: its very low-redshift ($z=0.042$) and low optical luminosity ($\log \nu_oL_{\nu_o}<43$~erg~s$^{-1}$) suggest it is not an AGN.

\subsection{Absorbed AGN}
The weak emission lines from SDSS J1511+5637 and SDSS J1658+6118 might be explained by absorption.  Both objects have extremely red spectra (optical spectral indices $\alpha_\nu=4.4$ and 5.9, respectively), and SDSS J1658+6118 is additionally classified as a broad absorption line quasar in \citet{trump06}.  

\subsection{Misidentified Optical Counterparts to RASS X-ray Sources}
There are sometimes multiple optical sources within each 60$''$ RASS X-ray circle, and we believe two radio-quiet \bl\ candidates (SDSS 0755+3525 and SDSS J1610+2536) are not the proper optical counterparts.    SDSS 0755+3525  is 48$''$ from the center of the RASS error circle.  There is a more likely optical counterpart (identified as a starburst galaxy by \citealt{brinkmann00})  just 3$''$ away from the RASS source and 0.2$''$ from a FIRST radio source (FIRST J075551.3+352635).  For SDSS J1610+2536, there is a bright star 16$''$ away (cataloged as V1024~Her, an eclipsing variable star with $V=12.5$~mag in SIMBAD) that is also inside the RASS X-ray circle.  V1024~Her is probably the X-ray emitter.  Although its spectral type is not available in the literature (and it is too bright to have an SDSS spectrum), its optical magnitude and RASS X-ray count rate are consistent with those expected from a K or M star (see Figure~3 of \citealt{agueros09}).  We remove both sources from our list of radio-quiet \bl\ candidates.

\section{Discussion}
\label{sec:disc}

\begin{figure*}
\centering
\includegraphics[scale=0.8]{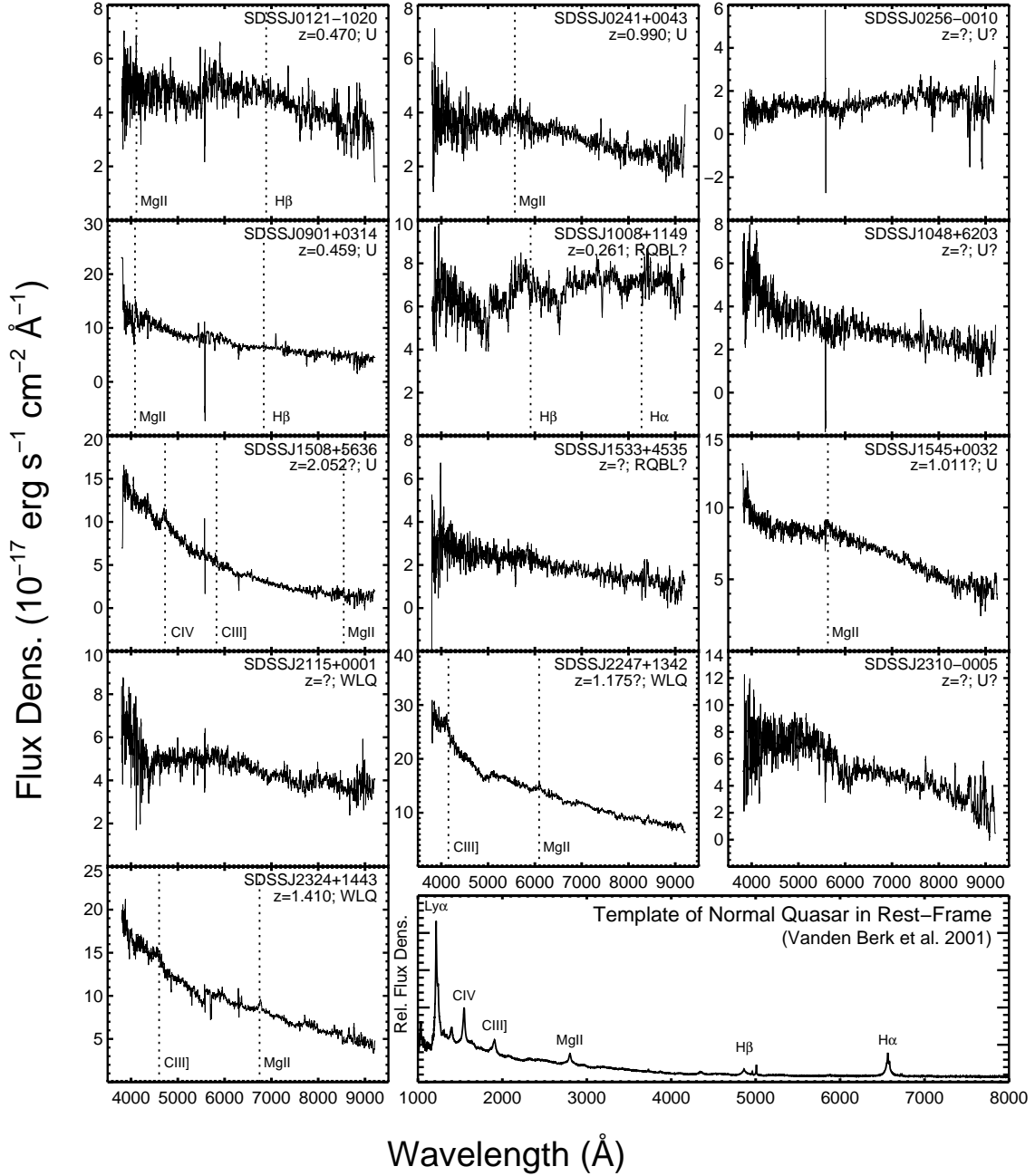} 
\caption{SDSS spectra of the 13 surviving weak-featured AGN in the observed frame.   In the bottom-right panel we show, for reference, the \citet{vandenberk01} composite SDSS quasar spectral template (shown in the rest-frame), with prominent broad emission lines labeled.  The expected locations of these lines are drawn for the radio-quiet \bl\ candidates with redshifts.   Redshifts smaller than 0.5 are derived from host galaxy absorption features (not labeled).    Our classification as Weak Line Quasar (WLQ), Radio-quiet \bl\ Candidate (RQBL?), or Unknown (U) from Table~\ref{tab:ch5_rquiet} is given after each object's redshift.   Some panels show a residual sky line at 5500~\AA. }
\label{fig:sampspec}
\end{figure*}

We are left with 13 AGN with weak emission features (see Figure~\ref{fig:sampspec}), 10 of which are confirmed to be radio-quiet from our VLA observations.  The other three sources' FIRST radio limits are not sensitive enough to determine if they are radio-quiet.   Our 13 surviving weak-featured AGN are likely a heterogeneous combination of at least two populations of objects.  Here, we classify them as likely low-redshift WLQs, as radio-faint \bl\ objects, or as uncertain.   These classifications are listed in Table~\ref{tab:ch5_rquiet}.  We also compare them to other low- and high-redshift radio-quiet AGN in the literature  with weak or absent spectral features.

\subsection{Low-redshift WLQs}
\label{sec:disc_wlq}

 \begin{figure*}
\centering
\includegraphics[scale=0.6]{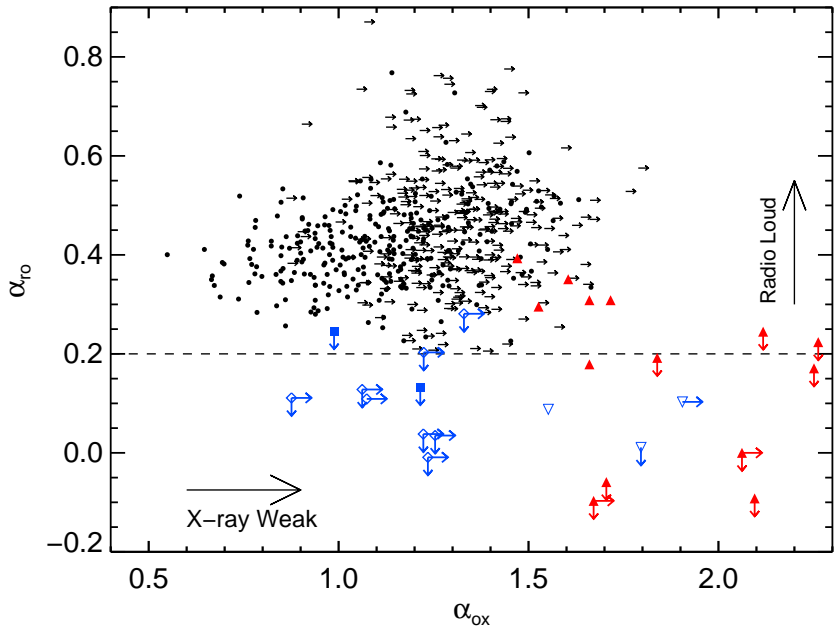}  
\caption{Broad-band spectral indices of 13 surviving $z<2.2$ radio-quiet \bl\ candidates, with arrows denoting limits.  {\it Filled blue squares:} two objects with RASS X-ray detections retained (with low-confidence) as potential very radio-faint \bl\ objects.  {\it Open blue diamonds}: eight objects lacking sensitive X-ray constraints.  {\it Open upside-down blue triangles:} three objects with {\it Chandra} X-ray observations, these are likely low-redshift WLQs.  For reference, we show 14 high-redshift ($2.7<z<5.9$) WLQs (filled red triangles) with Chandra/XMM X-ray follow-up from \citet{shemmer06} and \citet{shemmer09}.  Their radio information is typically derived from FIRST/NVSS, except for three WLQs with follow-up VLA observations in Table~\ref{tab:ch5_vla_wlq}.   We also plot $\sim$600 optically selected radio-loud \bl\ objects.  {\it Circles} denote \bl\ objects with an X-ray detection in RASS, and {\it arrows} mark X-ray limits for objects lacking RASS detections.    The dashed line  shows the traditional boundary between radio-loud/quiet quasars ($\alpha_{ro}=0.2$).}
\label{fig:ch5_rqalpha}
\end{figure*}

The multiwavelength colors of our 13 radio-quiet \bl\ candidates are shown in Figure~\ref{fig:ch5_rqalpha}.   The three objects with {\it Chandra} coverage (SDSS J2115+0001, SDSS J2247+1342, and SDSS J2324+1443, open upside down blue triangles) are radio and X-ray fainter than normal \bl\ objects.   SDSS J2115+0001 does not have a redshift, but it does have an 8.4~GHz radio detection implying $\alpha_{ro}=0.1$.  It also shows weak variability in its Stripe~82 light curve.   \citet{smith07} performed a polarimetric survey of 42 \bl\ candidates from \citetalias{collinge05}, and they did not detect strong polarization from either SDSS J2247+1342 or SDSS J2324+1443 (both have polarization $P<1\%$).  This further suggests they should not be unified with \bl\ objects.  Based on their low polarization and  multiwavelength colors, they may instead be low-redshift analogs to WLQs.

The prototype high-redshift WLQ was discovered by \citet[][SDSS J1532-0039; $z=4.62$]{fan99}, who remarked on a ``high-redshift quasar without emission lines.''  SDSS J1532-0039 was not detected in the X-ray with follow-on {\it Chandra} observations \citep{shemmer06}; it also remained undetected by the VLA  (clearly placing it as radio-quiet), and optical observations also did not find any polarization or strong variability \citep{diamond09}. \citet{anderson01} discovered two more $z=4.5-4.6$ WLQs, and a fourth discovery at $z=5.9$ was reported by \citet{fan06}.  \citet{diamond09} present a sample of $\sim$70 SDSS WLQs ($z>3$), defined by Ly$\alpha$+\ion{N}{5} $REW<15$~\AA.   They show that WLQs constitute a prominent excess of objects in the low Ly$\alpha$+\ion{N}{5} equivalent width tail of the high-redshift quasar  distribution (which otherwise  follows a log-normal distribution), and there is no corresponding excess of objects in the high-$REW$ tail.

While the SDSS has produced a relatively large number of high-redshift discoveries, there are only a handful of low-redshift candidates in the literature.  To our knowledge, PG 1407+265 \citep[$z=0.94$;][]{mcdowell95} was the first AGN with weak emission features not interpreted  as a \bl\ phenomenon. PG 1407+265 has radio, X-ray, and optical flux ratios typical of normal radio-quiet quasars. \citet{mcdowell95} present UV, optical, and near-infrared spectra, and they find all lines (in particular, Ly$\alpha$, \ion{C}{4} and \ion{Mg}{2}) have unusually small equivalent widths.  However, they do detect comparatively strong H$\alpha$ by this odd source's standards ($REW=126$~\AA, which is still relatively weak compared to most quasars).  \citet{blundell03} report the detection of a relativistically boosted core for PG 1407+265, which might (help) explain its odd spectral characteristics.  Even with the detection of a radio core, \citet{blundell03} do not believe PG 1407+265 has been radio loud within the last 10$^6$-10$^7$ years.

\citet{leighly07ii} present detailed multiwavelength spectral coverage of PHL 1811 ($z$=0.192), an unusual X-ray weak narrow-line type~1 quasar.  They detect relatively strong H$\alpha$ and H$\beta$ emission, but forbidden, semiforbidden, and high ionization lines are extraordinarily weak or absent.  PHL 1811 also shows relatively strong \ion{Fe}{2} and \ion{Fe}{3} emission.  We note in particular that the SDSS spectra of SDSS J2247+1342 and SDSS J2324+1442 look similar to PHL 1811, and we might expect them to also have strong \ion{Fe}{2} and \ion{Fe}{3} lines.

 \citet{leighly07ii} run photoionziation models and find almost all of the odd spectral characteristics of PHL 1811 can be explained by its soft  SED, as fewer high-ionization species are formed than would result from a harder (and more normal) SED.     Most of the radiative cooling is performed via Hydrogen lines (which can be excited by the soft continuum), hence the relatively strong low-order Balmer lines.    \citet{leighly07i} show PHL 1811 has a steep X-ray photon spectral index ($\Gamma\sim2.3$ over 0.3-5.0~keV), and they attribute this to an exceptionally high (perhaps super-Eddington) accretion rate.  \citet{shemmer09} note that {\it Chandra} detected high-redshift SDSS WLQs  do not appear to have unusually steep hard X-ray spectra,  but they stress that X-ray observations with improved photon statistics are required to test this properly.  Infrared spectroscopy to determine the strength of H$\alpha$ and/or H$\beta$ for SDSS WLQs would also be illuminating, and H$\beta$ line widths may allow accretion rate estimates \citep[e.g., see][]{shemmer04}.

A proposed lower-redshift analog to high-redshift WLQs was serendipitously discovered in the SDSS spectroscopic database by \citet[][SDSS J094533.99+100950.1; $z=1.66$]{hryniewicz10}.  SDSS J0945+1009 has a normal optical continuum compared to radio-quiet quasars, but it shows very weak \ion{Mg}{2} emission ($REW\sim15$~\AA) and absent \ion{C}{3}] and \ion{C}{4} emission.  SDSS J0945+1009 is similar to the intermediate redshift radio-quiet \bl\ candidates presented in Table~\ref{tab:ch5_rquiet}, however, even its weak \ion{Mg}{2} is too large to make it into our SDSS \bl\ samples.

\subsection{Radio-Quiet \bl\ Candidates}
\label{sec:disc_rqbl}
Two objects (SDSS J1008+1149 and SDSS J1533+4535) are detected in the X-ray by RASS, with X-ray brightnesses typical of \bl\ objects ($\alpha_{ox}$=0.99 and 1.22, respectively, see Figure~\ref{fig:ch5_rqalpha}).  Both objects appear to be the best optical counterparts in each RASS error circle, but of course {\it Chandra} observations would be helpful for confirmation.  These sources are very interesting, but we cannot currently assert they are extraordinary radio-quiet \bl\ objects:  we do not have deeper VLA constraints for SDSS 1008+1149, and its FIRST flux limit ($\alpha_{ro}<0.25$) does not firmly identify it as radio-quiet.  Its redshift ($z=0.26$) is also very typical of the parent SDSS \bl\ sample, and, similar to SDSS J0241+0009, this could simply be a relatively radio-faint but otherwise normal high-energy peaked \bl\ object.    SDSS J1533+4535 was not detected in deeper VLA observations, and it is radio-fainter than normal \bl\ objects at the $>$3.4$\sigma$ level.  However, its SDSS spectrum is relatively noisy.  Both sources should be placed at high-priority for further study.  

\subsection{Unknown}
\label{sec:disc_unknown}
The remaining eight objects are very likely AGN, but their RASS X-ray limits are not sensitive enough to determine if they are X-ray weaker than \bl\ objects (see Figure~\ref{fig:ch5_rqalpha}).  Deeper X-ray observations are necessary to determine if they are best unified with \bl\ objects or with WLQs.  We only lack deeper VLA observations for two objects -- SDSS J0901+0314 and SDSS J1048+6203.  

Three objects (SDSS J0241+0043, SDSS J1508+5636, and SDSS J1545+0032) are particularly difficult to unify with \bl\ objects because they are at relatively high redshifts: they all have $z>1$, and the median redshift of radio-loud SDSS \bl\ objects from \citetalias{plotkin10} with reliable redshifts is 0.34.    Polarimetric observations by \citet{smith07} also found SDSS J1545+0032 to not be highly polarized ($P\sim$1\%).  All three objects have $\alpha_{ro}^{vla}<0.2$, and we suggest they are most likely lower-redshift WLQs (although we do so with low-confidence given their poor X-ray constraints).

Two objects, SDSS J0121$-$1020 and SDSS J0901+0314, have redshifts ($z=0.47$ and $z=0.46$, respectively) typical of normal \bl\ objects.  SDSS J0121-1020 is definitively radio-quiet ($\alpha_{ro}<0.04$, $>$4.75$\sigma$ radio fainter than normal \bl\ objects); we do not have deeper VLA observations for SDSS J0901+0314, but its FIRST radio flux limit ($\alpha_{ro}^{first}<0.20$) indicates it is radio-fainter than \bl\ objects at the $>$2.75$\sigma$ level.  Deeper X-ray observations should be taken for both sources to determine if they are more likely very low-redshift WLQs or if they are very radio-faint \bl\ objects. 

Neither source shows obvious signatures of star formation or for obscured emission lines, and their \ion{Ca}{2}~H/K depressions are much smaller than expected for normal elliptical galaxies.  We do note that weak stellar absorption lines are seen in the spectrum of SDSS J0901+0314, which suggests it might be a post-starburst galaxy.  However, an enhanced blue continuum from recent star formation alone is unlikely strong enough to explain its weak \ion{Ca}{2}~H/K break and lack of strong emission lines.  For example, SDSS J0901+0314 shows only very weak H$\delta$ with $REW$ of 1--2~\AA\ in absorption, and post-starburst galaxies typically require H$\delta$ stronger than 5~\AA\ in absorption \citep[e.g., see][]{goto05}.   

 Interestingly, these two objects appear remarkably similar to  the radio-quiet \bl\ candidate 2QZ J215454.3$-$305654 ($z=0.49$) discovered by \citet{londish04};  2QZ J2154$-$3056 shows only weak [\ion{O}{3}] and perhaps some very weak Balmer emission.\footnote{H$\alpha$ is not in the spectral coverage of their optical spectrum.  \ion{Ca}{2}~H/K and G-band absorption from the host galaxy are also detected.}  
Both SDSS J0121$-$1020 and SDSS J0901+0314 also show weak forbidden lines from Oxygen and very weak narrow H$\beta$ (all lines have $REW<5$~\AA).   \citet{londish04} do not detect optical polarization or radio emission  ($\alpha_{ro}<0.08$), arguing against a beamed synchrotron continuum for 2QZ J2154$-$3056.  X-ray emission is not seen in RASS ($f_X<10^{-14}$~erg~s$^{-1}$~cm$^{-2}$), and we estimate from their Figure 2 that $\alpha_{ox}>1.2$.  They do detect marginal variability; the variability, however, is seen in only one of four epochs over a  $\sim$2-3 month period with a reduced $\chi_r^2=4$.  The lack of relatively strong H$\beta$, and its redder spectrum, makes 2QZJ 2154$-$3056 dissimilar to PHL 1811 and PG 1407+265.  

2QZJ 2154$-$3056 and the similar SDSS  sources are unlikely explained by near- or super-Eddington accretion rates (as in PHL~1811), because then we would expect relatively strong H$\beta$.  It is alternatively possible to create lineless AGN with exceptionally low accretion rates ($\lesssim$10$^{-2}$--10$^{-3}$~$\dot{M}_{Edd}$), as has been invoked for so called ``naked AGN'' (i.e., Type~1 Seyfert galaxies lacking broad emission lines; \citealt{hawkins04}) and X-ray bright, optically normal galaxies (XBONGS; e.g., see \citealt{trump09}).  At these low accretion rates, the inner region of the accretion disk becomes radiatively inefficient.  Thus, a radiation pressure driven wind never forms the BELR \citep[e.g., see][]{nicastro00,nicastro03}.  \citet{elitzur09} reach a similar conclusion based on mass conservation arguments. 

SDSS J0121$-$1020 and SDSS J0901+0314 are, however, unlikely to have low-enough accretion rates for their BELRs to vanish.  Even if they have extremely massive black holes (10$^{10}~M_{\sun}$), \citet{elitzur09} predict they would require radiatively inefficient accretion flows and  bolometric luminosities $L_{bol}<5\times10^{41}$~erg~s$^{-1}$ for their BELRs to disappear; this scenario is excluded by their optical luminosites.   Also, in the absence of a radiatively driven wind, the outflow is likely replaced by a jet, so we would expect very low accretion rate systems to also have strong radio emission \citep[e.g., see][]{ho08}.

It would be surprising if SDSS J0121$-$1020, SDSS J0901+0314, and 2QZJ 2154$-$3056 were relatively radio-faint but otherwise normal \bl\ objects with their emission lines simply diluted by starlight from the host galaxy.  \citet{marcha96} argue that the strongest emission features in weakly beamed \bl\ objects show equivalent widths up to $\sim$50-60~\AA\ when the equivalent width measures  are referenced to the contaminating starlight continuum.  In those cases, a boosted synchrotron continuum still acts to weaken the lines slightly (albeit less than if the AGN was more highly beamed).  We do not believe the host galaxy on its own could render H$\alpha$ or H$\beta$ completely undetectable.  In the absence of a boosted continuum, H$\alpha$ and H$\beta$ in these radio-quiet \bl\ candidates are likely intrinsically weak, but the mechanism is currently unclear (but see \citealt{londish04} for potential explanations).

Finally, three other objects (SDSS J0256$-$0010, SDSS J1048+6203, and SDSS J2310$-$0005) remain as weak-featured AGN, but we retain them with lower-confidence because they have relatively noisy SDSS spectra.  All three also lack spectroscopic redshifts, but SDSS J0256$-$0010 and SDSS J2310$-$0005 show variability in their Stripe~82 light curves.  Both of these sources are also confirmed to be radio-quiet in their deeper VLA observations.  We do not have VLA data for SDSS J1048+6203, and its FIRST limit is not terribly sensitive ($\alpha_{ro}<0.281$).  It is possible this source is a normal radio-loud \bl\ object, but deeper radio and X-ray observations are necessary to settle this issue.

\subsection{Radio-Quiet BL Lac Candidates from \citetalias{plotkin10}}  
There are 71 additional $z<2.2$ objects identified as radio-quiet \bl\ candidates in \citetalias{plotkin10}.  All have FIRST/NVSS radio flux densities or limits in the radio-quiet regime ($\alpha_{ro}<0.2$), so they are not just optically faint \bl\ objects with radio fluxes below the FIRST/NVSS limits.  All but five of their objects have extragalactic redshifts, so there appears to be little contamination from stars.   \citetalias{plotkin10} argue many of these objects are distinct from \bl\ objects because they tend to be at systematically higher redshifts and are more luminous than radio-loud \bl\ objects.  However, deeper X-ray observations are necessary: the majority (95\%) do not have RASS X-ray detections, and their RASS flux limits are not sensitive enough to discriminate them from \bl\ objects in the X-ray.    Based on our conclusions from our analysis of the \citetalias{collinge05}/\citetalias{anderson07} sample, we expect deeper X-ray observations will show the majority of \citetalias{plotkin10} objects to be low-redshift WLQs.

\section{Conclusions}
\label{sec:conc}
Here, we present follow-up radio and X-ray observations for a subset of $z<2.2$ radio-quiet \bl\ candidates from a sample of \nrqsamp\ sources originally presented in \citetalias{collinge05} and \citetalias{anderson07}.   When available, we also consider flux variability and proper motion measures, and we conclude that 13 of the 26 objects are AGN with weak-emission lines.  All but three of these 13 are definitively radio-fainter than \bl\ objects at the $\ge$2.75$\sigma$ level (the other three objects are not detected in the radio, but their radio flux limits are not sensitive enough to declare them radio-quiet), making it difficult to interpret them as beamed radio galaxies.  We thus confirm that the SDSS has recovered a population of radio-quiet AGN with intrinsically weak emission lines.   That is, relativistic beaming does not appear to be the dominant cause of these objects' featureless spectra (but we also cannot exclude the possibility that weak jets exist in some of these sources).  These 13 weak-featured AGN are likely a heterogenous mix of at least two populations of objects.

Of the 13 AGN, three with deeper {\it Chandra} X-ray observations are likely low-redshift WLQs, analogs to the $\sim$10$^2$ object high-redshift ($z>2.2$) population discovered by the SDSS.  Other low-redshift WLQ candidates have been discovered serendipitously in the SDSS \citep[e.g.,][]{hryniewicz10}, but the \citetalias{collinge05}/\citetalias{anderson07} \bl\ surveys were the first systematic searches of the SDSS database capable of recovering low-redshift WLQs.   

There are two radio-quiet \bl\ candidates that have X-ray detections in RASS and are too X-ray bright to be unified with WLQs, but their $\alpha_{ox}^{rass}$ measures are similar to those of \bl\ objects.  We retain these as radio-quiet \bl\ candidates, but with low-confidence.  One object's radio flux limit is not sensitive enough to declare it radio-quiet (i.e., it could be a relatively radio-faint but still normal radio-loud \bl\ object).  The other object is definitively radio-quiet, but its optical spectrum has relatively low $S/N$.  The fact that we do not find a single high-confidence radio-quiet \bl\ object supports the notion that all \bl\ objects are radio-loud \citep{stocke90}.  

There are eight other  \citetalias{collinge05}/\citetalias{anderson07} radio-quiet \bl\ candidates lacking RASS X-ray detections for which we do not have deeper {\it Chandra} coverage.  It is likely some of these will also be shown to be low-redshift WLQs upon more detailed X-ray scrutiny.  Many radio-quiet \bl\ candidates in \citetalias{plotkin10} are probably also WLQs, and we thus conclude that the SDSS has discovered a new population of low-redshift WLQs.
  
We compare our radio-quiet SDSS \bl\ candidates to other radio-quiet \bl\ candidates/WLQs in the literature.   We discuss how high-accretion rates can lead to soft ionizing continua and yield spectra with weak or absent UV lines but relatively strong H$\alpha$ and H$\beta$ (but those lines may still be weak compared to normal Type~1 quasars).  Near-infrared spectroscopy is required to test this properly.    Two of the eight objects lacking sensitive X-ray limits are similar to the radio-quiet \bl\ candidate discovered by \citet[][2QZJ2154-3056]{londish04}.  These objects are unlikely to have high mass accretion rates (because they do not show H$\beta$), and they are too luminous to have weak BELRs because of extraordinarily low accretion rates.

The radio-quiet \bl\ candidates should be monitored for polarization  to determine if these objects show signs of weakly beamed relativistic jets.  Objects with relatively larger (smaller) $\alpha_{ro}$ ($\alpha_{ox}$) values or limits  similar to those for radio-loud \bl\ objects should be placed at high priority for polarimetric monitoring.  Similarly, searches for flux variability and deeper radio observations for objects with poorer $\alpha_{ro}$ constraints would be useful.   The detection of hot thermal dust emission in the infrared would strongly argue against a boosted continuum, as beaming would dilute  dust emission \citep[e.g., see][]{diamond09}. 

\citet{londish04} successfully model the spectrum of 2QZJ 2154$-$3056 as the combination of an elliptical galaxy, a starburst, and a central AGN, and an image decomposition shows a host galaxy and AGN.   Their optical image cannot discriminate between an elliptical or disk host.  Interestingly, PHL 1811 appears to live in a spiral galaxy \citep{jenkins05}.  \bl\ objects are hosted (perhaps exclusively) by giant elliptical galaxies \citep[e.g., see][]{urry00_hstii}.  Thus, host galaxy imaging, which could be achieved at least for the lowest-redshift radio-quiet \bl\ candidates, might be a useful diagnostic.

The SDSS has discovered a population of at least 13 low-redshift ($z<2.2$) radio-quiet AGN with weak-featured optical spectra, many of which we believe are low-redshift analogs to WLQs.   The \citetalias{plotkin10} optically selected \bl\ sample adds up to another 71 objects.  The case of SDSS J0945+1009 \citep{hryniewicz10} with weak \ion{Mg}{2} $REW \sim 15$~\AA\ suggests there are likely even more objects yet to be recognized in the SDSS database (as SDSS \bl\ searches exclude objects with any emission feature larger than 5~\AA\ rest-frame).   In all likelihood, the SDSS radio-quiet \bl\ candidates are a heterogeneous mix of at least two populations of objects, so it is going to take even more observational effort to disentangle the physical nature of these strange objects.

\acknowledgements
We would like to thank Adrienne Stilp for her assistance reducing the VLA data, and we also thank the anonymous referee for helpful comments to improve this manuscript.  This research has made use of software provided by the Chandra X-ray Center (CXC) in the application packages CIAO.   R.M.P.\ and S.F.A.\ gratefully acknowledge support from NASA/ADP grant NNG05GC45G, and  from  the National Aeronautics and Space Administration through Chandra Award Number GO9-0126X issued by the {\it Chandra X-ray Observatory} Center, which is operated by the Smithsonian Astrophysical Observatory for and on behalf of the National Aeronautics Space Administration under contract NAS8-3060.  W.N.B.\ acknowledges support from NASA/ADP grant NNX10AC99G.  

Funding for the SDSS and SDSS-II has been provided by the Alfred P. Sloan Foundation, the Participating Institutions, the National Science Foundation, the U.S. Department of Energy, the National Aeronautics and Space Administration, the Japanese Monbukagakusho, the Max Planck Society, and the Higher Education Funding Council for England. The SDSS Web Site is \url{http://www.sdss.org/.}

The SDSS is managed by the Astrophysical Research Consortium for the Participating Institutions. The Participating Institutions are the American Museum of Natural History, Astrophysical Institute Potsdam, University of Basel, University of Cambridge, Case Western Reserve University, University of Chicago, Drexel University, Fermilab, the Institute for Advanced Study, the Japan Participation Group, Johns Hopkins University, the Joint Institute for Nuclear Astrophysics, the Kavli Institute for Particle Astrophysics and Cosmology, the Korean Scientist Group, the Chinese Academy of Sciences (LAMOST), Los Alamos National Laboratory, the Max-Planck-Institute for Astronomy (MPIA), the Max-Planck-Institute for Astrophysics (MPA), New Mexico State University, Ohio State University, University of Pittsburgh, University of Portsmouth, Princeton University, the United States Naval Observatory, and the University of Washington.



\begin{thebibliography}{64}
\expandafter\ifx\csname natexlab\endcsname\relax\def\natexlab#1{#1}\fi

\bibitem[{Abazajian {et~al.}(2009)}]{dr7pap}
Abazajian, K. {et~al.} 2009, \apjs, 182, 543

\bibitem[{Ag{\"u}eros {et~al.}(2009)Ag{\"u}eros, Anderson, Covey,
  {et~al.}}]{agueros09}
Ag{\"u}eros, M.~A., {et~al.} 2009, \apjs, 181,
  444

\bibitem[{Aller {et~al.}(1999)Aller, Aller, Hughes, \& Latimer}]{aller99}
Aller, M.~F., Aller, H.~D., Hughes, P.~A., \& Latimer, G.~E. 1999, \apj, 512,
  601

\bibitem[{Anderson {et~al.}(2001)Anderson, Fan, Richards,
  {et~al.}}]{anderson01}
Anderson, S.~F.,  {et~al.} 2001, \aj, 122, 503

\bibitem[{Anderson {et~al.}(2007)Anderson, Margon, Voges,
  {et~al.}}]{anderson07}
Anderson, S.~F., {et~al.} 2007, \aj, 133, 313 (A07)

\bibitem[{Becker {et~al.}(1995)Becker, White, \& Helfand}]{becker95}
Becker, R.~H., White, R.~L., \& Helfand, D.~J. 1995, \apj, 450, 559

\bibitem[{Blandford \& Rees(1978)}]{blandford78}
Blandford, R.~D. \& Rees, M.~J. 1978, in Wolfe A.M., ed., Pitt.~Conf.~on BL~Lac
  Obj., Pittsburgh, p.~328

\bibitem[{Blundell {et~al.}(2003)Blundell, Beasley, \& Bicknell}]{blundell03}
Blundell, K.~M., Beasley, A.~J., \& Bicknell, G.~V. 2003, \apjl, 591, L103

\bibitem[{Brinkmann {et~al.}(2000)Brinkmann, Laurent-Muehleisen, Voges,
  Siebert, Becker, Brotherton, White, \& Gregg}]{brinkmann00}
Brinkmann, W., Laurent-Muehleisen, S.~A., Voges, W., Siebert, J., Becker,
  R.~H., Brotherton, M.~S., White, R.~L., \& Gregg, M.~D. 2000, \aap, 356, 445

\bibitem[{Collinge {et~al.}(2005)Collinge, Strauss, Hall,
  {et~al.}}]{collinge05}
Collinge, M.~J., {et~al.} 2005, \aj, 129, 2542 (C05)

\bibitem[{Condon {et~al.}(1998)Condon, Cotton, Greisen, Yin, Perley, Taylor, \&
  Broderick}]{condon98}
Condon, J.~J., Cotton, W.~D., Greisen, E.~W., Yin, Q.~F., Perley, R.~A.,
  Taylor, G.~B., \& Broderick, J.~J. 1998, \aj, 115, 1693

\bibitem[{Diamond-Stanic {et~al.}(2009)}]{diamond09}
Diamond-Stanic, A.~M. {et~al.} 2009, \apj, 699, 782

\bibitem[{Elitzur \& Ho(2009)}]{elitzur09}
Elitzur, M. \& Ho, L.~C. 2009, \apj, 701, L91

\bibitem[{Fan {et~al.}(2006)Fan, Strauss, Richards, {et~al.}}]{fan06}
Fan, X.,  {et~al.} 2006, \aj, 131, 1203

\bibitem[{Fan {et~al.}(1999)}]{fan99}
Fan, X. {et~al.} 1999, \apjl, 526, L57

\bibitem[{Garmire {et~al.}(2003)Garmire, Bautz, Ford, Nousek, \&
  Ricker}]{garmire03}
Garmire, G.~P., Bautz, M.~W., Ford, P.~G., Nousek, J.~A., \& Ricker, G.~R.
  2003, Proc.\ SPIE, 4851, 28

\bibitem[{Gehrels(1986)}]{gehrels86}
Gehrels, N. 1986, \apj, 303, 336

\bibitem[{Goto(2005)}]{goto05} Goto, T.\ 2005, \mnras, 357, 937 

\bibitem[{Hawkins(2004)}]{hawkins04}
Hawkins, M.~R.~S. 2004, \aap, 424, 519

\bibitem[{Ho(2008)}]{ho08} Ho, L.~C.\ 2008, \araa, 46, 475 

\bibitem[{Hryniewicz {et~al.}(2010)Hryniewicz, Czerny, Nikolajuk, \&
  Kuraszkiewicz}]{hryniewicz10}
Hryniewicz, K., Czerny, B., Nikolajuk, M., \& Kuraszkiewicz, J. 2010, \mnras,
 404, 2028

\bibitem[{Ivezi{\'c} {et~al.}(2007)Ivezi{\'c}, Smith, Miknaitis,
  {et~al.}}]{ivezic07}
Ivezi{\'c}, {\v Z}., Smith, J.~A., Miknaitis, G., {et~al.} 2007, \aj, 134, 973

\bibitem[{Jannuzi {et~al.}(1993)Jannuzi, Green, \& French}]{jannuzi93}
Jannuzi, B.~T., Green, R.~F., \& French, H. 1993, \apj, 404, 100

\bibitem[Jenkins {et~al.}(2005)]{jenkins05} Jenkins, E.~B., Bowen, 
D.~V., Tripp, T.~M., \& Sembach, K.~R.\ 2005, \apj, 623, 767 

\bibitem[{Just {et al.}(2007)}]{just07} Just, D.~W., Brandt, 
W.~N., Shemmer, O., Steffen, A.~T., Schneider, D.~P., Chartas, G., 
\& Garmire, G.~P.\ 2007, \apj, 665, 1004 

\bibitem[{Kellermann {et~al.}(1989)Kellermann, Sramek, Schmidt, Shaffer, \&
  Green}]{kellermann89}
Kellermann, K.~I., Sramek, R., Schmidt, M., Shaffer, D.~B., \& Green, R. 1989,
  \aj, 98, 1195

\bibitem[{Kelly {et~al.}(2009)Kelly, Bechtold, \& Siemiginowska}]{kelly09}
Kelly, B.~C., Bechtold, J., \& Siemiginowska, A. 2009, \apj, 698, 895

\bibitem[{Kollgaard(1994)}]{kollgaard94}
Kollgaard, R.~I. 1994, Vistas in Astronomy, 38, 29

\bibitem[{Landt {et al.}(2002)}]{landt02}
 Landt, H., Padovani, P., \& Giommi, P.\ 2002, \mnras, 336, 945 

\bibitem[{Leighly {et~al.}(2007{\natexlab{a}})Leighly, Halpern, Jenkins, \&
  Casebeer}]{leighly07ii}
Leighly, K.~M., Halpern, J.~P., Jenkins, E.~B., \& Casebeer, D.
  2007{\natexlab{a}}, \apjs, 173, 1

\bibitem[{Leighly {et~al.}(2007{\natexlab{b}})Leighly, Halpern, Jenkins, Grupe,
  Choi, \& Prescott}]{leighly07i}
Leighly, K.~M., Halpern, J.~P., Jenkins, E.~B., Grupe, D., Choi, J., \&
  Prescott, K.~B. 2007{\natexlab{b}}, \apj, 663, 103

\bibitem[{Londish {et~al.}(2004)Londish, Heidt, Boyle, Croom, \&
  Kedziora-Chudczer}]{londish04}
Londish, D., Heidt, J., Boyle, B.~J., Croom, S.~M., \& Kedziora-Chudczer, L.
  2004, \mnras, 352, 903
  
\bibitem[{MacLeod {et~al.}(2010)}]{macleod10_ph} MacLeod, C.~L., et al.\ 
2010, \apj, subm., arXiv:1004.0276 

\bibitem[{Mannucci {et~al.}(2001)Mannucci, Basile, Poggianti, Cimatti, Daddi,
  Pozzetti, \& Vanzi}]{mannucci01}
Mannucci, F., Basile, F., Poggianti, B.~M., Cimatti, A., Daddi, E., Pozzetti,
  L., \& Vanzi, L. 2001, \mnras, 326, 745

\bibitem[{March{\~a} {et~al.}(1996)March{\~a}, Browne, Impey, \&
  Smith}]{marcha96}
March{\~a}, M., Browne, I.~W.~A., Impey, C.~D., \& Smith, P.~S. 1996, \mnras,
  281, 425

\bibitem[{Marscher \& Gear(1985)}]{marscher85}
Marscher, A.~P. \& Gear, W.~K. 1985, \apj, 298, 114

\bibitem[{Massaro {et~al.}(2009)Massaro, Giommi, Leto, Marchegiani, Maselli,
  Perri, Piranomonte, \& Sclavi}]{massaro09}
Massaro, E., Giommi, P., Leto, C., Marchegiani, P., Maselli, A., Perri, M.,
  Piranomonte, S., \& Sclavi, S. 2009, \aap, 495, 691

\bibitem[{McDowell {et~al.}(1995)McDowell, Canizares, Elvis, Lawrence, Markoff,
  Mathur, \& Wilkes}]{mcdowell95}
McDowell, J.~C., Canizares, C., Elvis, M., Lawrence, A., Markoff, S., Mathur,
  S., \& Wilkes, B.~J. 1995, \apj, 450, 585

\bibitem[{Mukai(1993)}]{mukai93}
Mukai, K. 1993, Legacy, vol.~3, p.21-31, 3, 21

\bibitem[{Munn {et~al.}(2004)Munn, Monet, Levine, {et~al.}}]{munn04}
Munn, J.~A. {et~al.} 2004, \aj, 127, 3034

\bibitem[{Nicastro(2000)}]{nicastro00}
Nicastro, F. 2000, \apjl, 530, L65

\bibitem[{Nicastro {et~al.}(2003)Nicastro, Martocchia, \& Matt}]{nicastro03}
Nicastro, F., Martocchia, A., \& Matt, G. 2003, \apjl, 589, L13

\bibitem[{Padovani \& Giommi(1995{\natexlab{a}})}]{padovani95_apj}
Padovani, P. \& Giommi, P. 1995{\natexlab{a}}, \apj, 444, 567

\bibitem[{Padovani \& Giommi(1995{\natexlab{b}})}]{padovani95_mnras}
---. 1995{\natexlab{b}}, \mnras, 277, 1477

\bibitem[{Perlman {et~al.}(2001)Perlman, Padovani, Landt, Stocke, Costamante,
  Rector, Giommi, \& Schachter}]{perlman01}
Perlman, E.~S., Padovani, P., Landt, H., Stocke, J.~T., Costamante, L., Rector,
  T., Giommi, P., \& Schachter, J.~F. 2001, in Blazar Demographics and Physics,
  227, 200

\bibitem[{Plotkin {et~al.}(2010)Plotkin, Anderson, Brandt,
  {et~al.}}]{plotkin10}
Plotkin, R.~M.,  {et~al.} 2010, \aj, 139, 390 (P10)

\bibitem[{Plotkin {et~al.}(2008)Plotkin, Anderson, Hall, Margon, Voges,
  Schneider, Stinson, \& York}]{plotkin08}
Plotkin, R.~M., Anderson, S.~F., Hall, P.~B., Margon, B., Voges, W., Schneider,
  D.~P., Stinson, G., \& York, D.~G. 2008, \aj, 135, 2453

\bibitem[{Schlegel {et~al.}(1998)Schlegel, Finkbeiner, \& Davis}]{schlegel98}
Schlegel, D.~J., Finkbeiner, D.~P., \& Davis, M. 1998, \apj, 500, 525

\bibitem[{Schneider {et~al.}(2007)Schneider, Hall, Richards,
  {et~al.}}]{schneider07}
Schneider, D.~P., Hall, P.~B., Richards, G.~T., {et~al.} 2007, \aj, 134, 102

\bibitem[{Schneider {et~al.}(2010)Schneider, Richards, Hall,
  {et~al.}}]{schneider10_ph}
Schneider, D.~P., Richards, G.~T., Hall, P.~B., {et~al.} 2010, AJ, 139, 2360
  
\bibitem[{Sesar {et~al.}(2007)Sesar, Ivezi{\'c}, Lupton, {et~al.}}]{sesar07}
Sesar, B., {et~al.} 2007, \aj, 134, 2236


\bibitem[{Shemmer {et~al.}(2009)Shemmer, Brandt, Anderson, Diamond-Stanic, Fan,
  Richards, Schneider, \& Strauss}]{shemmer09}
Shemmer, O., Brandt, W.~N., Anderson, S.~F., Diamond-Stanic, A.~M., Fan, X.,
  Richards, G.~T., Schneider, D.~P., \& Strauss, M.~A. 2009, \apj, 696, 580

\bibitem[{Shemmer {et~al.}(2008)Shemmer, Brandt, Netzer, Maiolino, \&
  Kaspi}]{shemmer08}
Shemmer, O., Brandt, W.~N., Netzer, H., Maiolino, R., \& Kaspi, S. 2008, \apj,
  682, 81

\bibitem[Shemmer {et~al.}(2004)]{shemmer04} Shemmer, O., Netzer, 
H., Maiolino, R., Oliva, E., Croom, S., Corbett, E., 
\& di Fabrizio, L.\ 2004, \apj, 614, 547 

\bibitem[{Shemmer {et~al.}(2006)}]{shemmer06}
Shemmer, O. {et~al.} 2006, \apj, 644, 86

\bibitem[{Smith {et~al.}(2007)Smith, Williams, Schmidt, Diamond-Stanic, \&
  Means}]{smith07}
Smith, P.~S., Williams, G.~G., Schmidt, G.~D., Diamond-Stanic, A.~M., \& Means,
  D.~L. 2007, \apj, 663, 118

\bibitem[{Stark {et~al.}(1992)}]{stark92}
Stark, A.~A. {et~al.} 1992, \apjs, 79, 77

\bibitem[{Steffen {et~al.}(2006)}]{steffen06} Steffen, A.~T., 
Strateva, I., Brandt, W.~N., Alexander, D.~M., Koekemoer, A.~M., Lehmer, 
B.~D., Schneider, D.~P., \& Vignali, C.\ 2006, \aj, 131, 2826 

\bibitem[{Stickel {et~al.}(1991)}]{stickel91} Stickel, M., Padovani, 
P., Urry, C.~M., Fried, J.~W., \& Kuehr, H.\ 1991, \apj, 374, 431 

\bibitem[{Stocke {et~al.}(1985)Stocke, Liebert, Schmidt, Gioia, Maccacaro,
  Schild, Maccagni, \& Arp}]{stocke85}
Stocke, J.~T., Liebert, J., Schmidt, G., Gioia, I.~M., Maccacaro, T., Schild,
  R.~E., Maccagni, D., \& Arp, H.~C. 1985, \apj, 298, 619

\bibitem[{Stocke {et~al.}(1992)Stocke, Morris, Weymann, \& Foltz}]{stocke92}
Stocke, J.~T., Morris, S.~L., Weymann, R.~J., \& Foltz, C.~B. 1992, \apj, 396,
  487

\bibitem[{Stocke {et~al.}(1990)}]{stocke90}
Stocke, J.~T., {et~al.} 1990, \apj, 348, 141

\bibitem[{Stocke {et~al.}(1991)}]{stocke91}
---. 1991, \apjs, 76, 813

\bibitem[{Tananbaum {et~al.}(1979)Tananbaum, Avni, Branduardi, Elvis, Fabbiano,
  Feigelson, Giacconi, Henry, Pye, Soltan, \& Zamorani}]{tananbaum79}
Tananbaum, H. et~al.  \apjl, 234, L9

\bibitem[{Trump {et~al.}(2009)Trump, Impey, Taniguchi, Brusa, Civano, Elvis,
  Gabor, Jahnke, Kelly, Koekemoer, Nagao, Salvato, Shioya, Capak, Huchra,
  Kartaltepe, Lanzuisi, McCarthy, Maineri, \& Scoville}]{trump09}
Trump, J.~R., {et~al.} \apj, 706, 797

\bibitem[{Trump {et~al.}(2006)}]{trump06}
Trump, J.~R., {et~al.} 2006, \apjs, 165, 1

\bibitem[{Ulrich {et~al.}(1997)Ulrich, Maraschi, \& Urry}]{ulrich97}
Ulrich, M.-H., Maraschi, L., \& Urry, C.~M. 1997, \araa, 35, 445

\bibitem[{Urry \& Padovani(1995)}]{urry95}
Urry, C.~M. \& Padovani, P. 1995, PASP, 107, 803

\bibitem[{Urry {et~al.}(2000)Urry, Scarpa, O'Dowd, Falomo, Pesce, \&
  Treves}]{urry00_hstii}
Urry, C.~M., Scarpa, R., O'Dowd, M., Falomo, R., Pesce, J.~E., \& Treves, A.
  2000, \apj, 532, 816

\bibitem[{Vanden Berk {et~al.}(2001)}]{vandenberk01} Vanden Berk, D.~E., 
et al.\ 2001, \aj, 122, 549 

\bibitem[{Voges {et~al.}(1999)Voges, Aschenbach, Boller, {et~al.}}]{voges99}
Voges, W. {et~al.} 1999, \aap, 349, 389

\bibitem[{Voges {et~al.}(2000)}]{voges00}
Voges, W. {et~al.} 2000, \iaucirc, 7432

\bibitem[{York {et~al.}(2000)York, Adelman, Anderson, {et~al.}}]{york00}
York, D.~G. {et~al.} 2000, \aj, 120, 1579

\end{thebibliography}
\end{document}